\documentstyle[lprocl,11pt,epsfig]{article}
\def\be{\begin{equation}}
\def\beq{\begin{equation}}
\def\ee{\end{equation}}
\def\eeq{\end{equation}}
\def\bea{\begin{eqnarray}}
\def\eea{\end{eqnarray}}
\newcommand{\beqa}{\begin{eqnarray}}
\newcommand{\eeqa}{\end{eqnarray}}

\newcommand{\bc}{\begin{center}}
\newcommand{\ec}{\end{center}}

 \def\(({\left(}
 \def\)){\right)}
\def\[[{\left[}
\def\]]{\right]}
\def\bi{\bibitem}

\def\a{\alpha}

\def\a{\alpha}
\def\b{\beta}
\def\g{\gamma}
\def\jabgd{J_{\alpha\beta\gamma\delta}}

\def\s{\sigma}
\begin{document}
\title{
Out of Equilibrium dynamics in Spin-Glasses and other Glassy Systems
}
\author{
Jean-Philippe Bouchaud}
\address{Service
de Physique de l'Etat Condens\'e, CEA-Saclay, Orme des Merisiers, 91 191 Gif
s/ Yvette CEDEX, France
}
\author{
Leticia F. Cugliandolo
}
\address{Laboratoire de Physique Th\'eorique de l'\'Ecole Normale Sup\'erieure$^a$,
24 rue Lhomond, 75231 Paris Cedex 05, France
}
\author{
Jorge Kurchan
}
\address{Laboratoire de Physique Th\'eorique de l'\'Ecole Normale Sup\'erieure
de Lyon,
46 All\'ee d'Italie, F-69364, Lyon Cedex-07, France
}
\author{
Marc M\'ezard
}
\address
{Laboratoire de Physique Th\'eorique de l'\'Ecole Normale Sup\'erieure
\footnote {Unit\'e propre du CNRS,  associ\'ee
 \`a\ l'Ecole
 Normale Sup\'erieure et \`a\ l'Universit\'e de Paris Sud} , 24 rue
 Lhomond, 75231 Paris Cedex 05, France
}
\vskip .2in
\maketitle

\vspace{1cm}

\abstracts{We review recent theoretical progress on glassy dynamics, with 
special emphasis on the importance and universality of the
{\it aging regime}, which is relevant to many experimental situations. The three main 
subjects which we address are: 
(i) Phenomenological models of aging (coarsening, trap models),  
(ii) Analytical results for the low-temperature dynamics of 
mean-field models (corresponding to the mode-coupling equations); and
(iii) Simple non-disordered models with glassy dynamics.
We discuss the interrelation between these approaches, and also with previous 
work in the field. Several open problems 
are underlined -- in particular the precise relation between mean-field like 
(or mode-coupling) descriptions and finite dimensional problems.}

\newpage

{\bf Table of Contents}
\vskip .2in
\vskip .2in

{\bf 1. Introduction}

\indent\indent {\it 1.1 Experiments: History dependence and Aging}

\indent\indent {\it 1.2 Think in the two times plane : time sectors.}

\indent\indent {\it 1.3 Fluctuation-Dissipation Relations}

\indent\indent {\it 1.4 Edwards-Anderson parameter, weak-ergodicity breaking
and
clonation}

{\bf 2. Phenomenological models of aging}

\indent\indent {\it 2.1 Coarsening in non disordered systems}

\indent\indent {\it 2.2 Coarsening in disordered systems I: Random ferromagnet
or random
field}

\indent\indent {\it 2.3 Coarsening in disordered systems II: Spin glasses and
droplets}

\indent\indent {\it 2.4 Stranded in phase space: the `trap' model}

{\bf 3. Mean-field models of aging: analytical results}

\indent\indent {\it 3.1 Self-averageness and universality}

\indent\indent {\it 3.2 The high temperature phase: two types of spin glasses}

\indent\indent {\it 3.3 Low temperatures: Weak long term memory and weak
ergodicity breaking}

\indent\indent {\it 3.4 Low temperature solution of the dynamical equations}

\indent\indent {\it 3.5 Generalization to several coupled modes --- the case of
spatial dependence}

\indent\indent {\it 3.6 Speculations on the effective age function $h(t)$}

\indent\indent {\it 3.7 Out of equilibrium versus `equilibrium on diverging
time scales'}

\indent\indent {\it 3.8 On the links between the static and dynamical
approaches.
Phase space geometry}


{\bf 4. Glasses and spin-glasses without disorder}

\indent\indent {\it 4.1 Phenomenology of glasses: a few basic facts}

 \indent\indent {\it 4.2 Discontinuous spin glasses: a mean-field scenario for
structural glasses}

\indent\indent {\it 4.3 Self-Induced Quenched Disorder: Spin glasses without
disorder}

 \indent\indent {\it 4.4 p-spin models, Mode Coupling Theory of glasses, and
its extension at low

\indent\indent \indent temperatures}

{\bf 5. Conclusion. Where do we stand ?}

\vskip 2cm

{ {\bf \footnotesize
LPTENS preprint 97/4}}

\vskip 2cm

\vspace{.5cm}
\section{Introduction}

Glassy systems are characterized by the fact that their relaxation time becomes
exceedingly long for low
temperatures, so long that these systems are never in equilibrium on laboratory
(or even geological) time
scales. Notwithstanding, most theories of spin glasses and other disordered
systems have first aimed at
describing the putative equilibrium state of these systems.
\cite{Edan}$^-$\cite{Fihe}
In doing so, many
difficulties and surprises
have emerged -- most notably the intricate structure of the Parisi solution of
the Sherrington-Kirkpatrick ({\sc sk}) \cite{Shki} mean-field model for
spin-glasses \cite{Pa}.
Despite some early attempts \cite{Sozi,So}, phenomenological and analytical
descriptions of the
non-equilibrium phenomena in disordered
systems have only recently appeared, which we shall review below. These {\it
dynamical}
approaches have been developed mostly because of the accumulating body of
experimental data on {\it aging} \cite{St}$^-$\cite{Nosvrev},
which is a striking
experimental consequence of the fact that these systems are out of equilibrium
even on macroscopic time scales. This aging regime is not the most
general out of equilibrium situation: a certain degree of
{\it universality} emerges in the
non-equilibrium properties.  While usual equilibrium dynamics is stationary,
i.e.
invariant under time translations, the aging regime presents a kind of
`covariance':
after transients have decayed,
the dynamical evolution of an old system of age $t_w$ is described by the
same equations as that of a younger system of age $t_w/2$, up to a rescaling
of time.

The change of focus from equilibrium to non equilibrium situations also has the
interesting consequence
of unveiling strong analogies between disordered systems such as spin-glasses,
and other types of
glasses where disorder is a priori absent, such as fragile glasses. These
analogies are both phenomenological
and formal: many experimental facts are in close correspondance (for example,
aging phenomena were first
studied in detail by Struick on polymer glasses \cite{St}), but also, the
structure of
the mean-field equations used to describe non-equilibrium spin-glasses
are almost
identical to the Mode Coupling
Theory ({\sc mct}) of supercooled liquids \cite{leut,Go}. There is thus a
strong feeling that the two
types of systems should
be deeply connected \cite{Kithwo}$^-$\cite{Kithwo2}, and there have been
several attempts in the past few years
to establish some precise bridges, which we shall review in this paper.

The scope of the present paper is mainly descriptive: we focus on general ideas
and concepts rather than on more technical aspects. We refer the reader to the
relevant papers for more details.

\subsection{Experiments: History dependence and Aging}
\label{experiments}

The simplest way to see that spin-glasses below the phase transition
temperature $T_g$ are not in equilibrium
even after times of the order of hours (or more) is the following: the sample
is quenched rapidly (under zero magnetic field) from high temperatures $T \gg
T_g$ to the working temperature $T_1<T_g$ which is reached, by convention, at
time
$t=0$. Then a very small oscillating field is applied to measure the a.c.
susceptibility $\chi$ of the sample at a certain frequency $\omega$.
What is observed is a slow continuous decrease
of the amplitude of
$\chi$ as a function of the time $t_w$ elapsed since the sample
reached the temperature $T_1$, which is called an {\it aging effect}.
\cite{expsuecia1}$^-$\cite{Nosvrev}
In
other words, $\chi$ is a function of both frequency {\it and} time:
$\chi(\omega,t_w)$. The response of the
system to a perturbation thus depends on the thermal history.

To a good approximation, the shape of $\chi(\omega,t_w)$ can be parametrized as
follows (see Fig.1):
\be
\chi(\omega,t_w) = A \; (\omega t_w)^{-b} + \chi_{\sc st}(\omega)
\qquad \qquad \qquad
(\omega t_w > 1) \label{chi}
\ee
where $A$ is a temperature dependent amplitude  and $b$ an
exponent that moves in the range $0.1 \to 0.4$.
The important points of the above parametrization are:

\begin{itemize}
\item
The response function is the sum of a {\it stationary part}
$\chi_{\sc st}(\omega)$
which
is independent of the age of the system $t_w$, and of an {\it aging} (or
non-stationary) contribution, which
decreases with time. $\chi_{\sc st}(\omega)$ behaves as $\omega^a$
with a small exponent $a$ (sometimes called $\alpha$)
\cite{expsaclay1,saclayrev1},
or perhaps as $\log \omega$. For
systems in equilibrium, the time dependent (aging) contribution
disappears.

\item
The aging contribution can be described with a
function of the scaling variable \cite{saclayrev1,saclayrev2}
$\omega t_w$. In general, the susceptibility of a system with a single
relaxation time
$\tau$ is a function of $\omega \tau$. Hence, the
above scaling form means that the effective relaxation time of the system is of
the order of its age $t_w$ itself. (See \cite{saclayrev2} for a discussion of
the
inset in Fig. 1 and a more detailed discussion of an alternative
description of the $\chi''_{\sc ag}$ data.)
\end{itemize}

\begin{figure}
\centerline{\epsfxsize=9cm
\epsffile{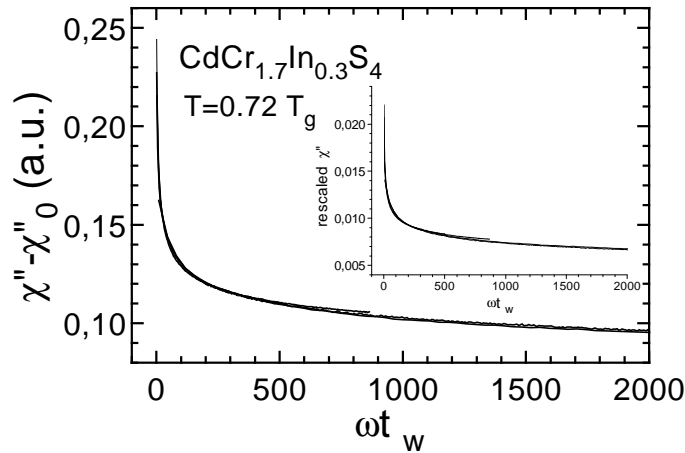}
}
\caption{The amplitude of the out of phase
magnetic
susceptibility
$\chi''(\omega,t_w)$ vs $\omega t_w$ for the
insulating spin-glass $CdCr_{1.7}In_{0.3}S_4$ (from Ref. [14]).
The  frequencies are $\omega=0.01, 0.03,0.1,1.$ Hz
and $t_w$ is the total time elapsed after the quench.
In the inset, a more refined scaling for $\chi''(\omega,t_w)$
as discussed in Ref.[14].}
\label{revfigchiexp}
\end{figure}

\begin{figure}
\centerline{\epsfxsize=10cm
\epsffile{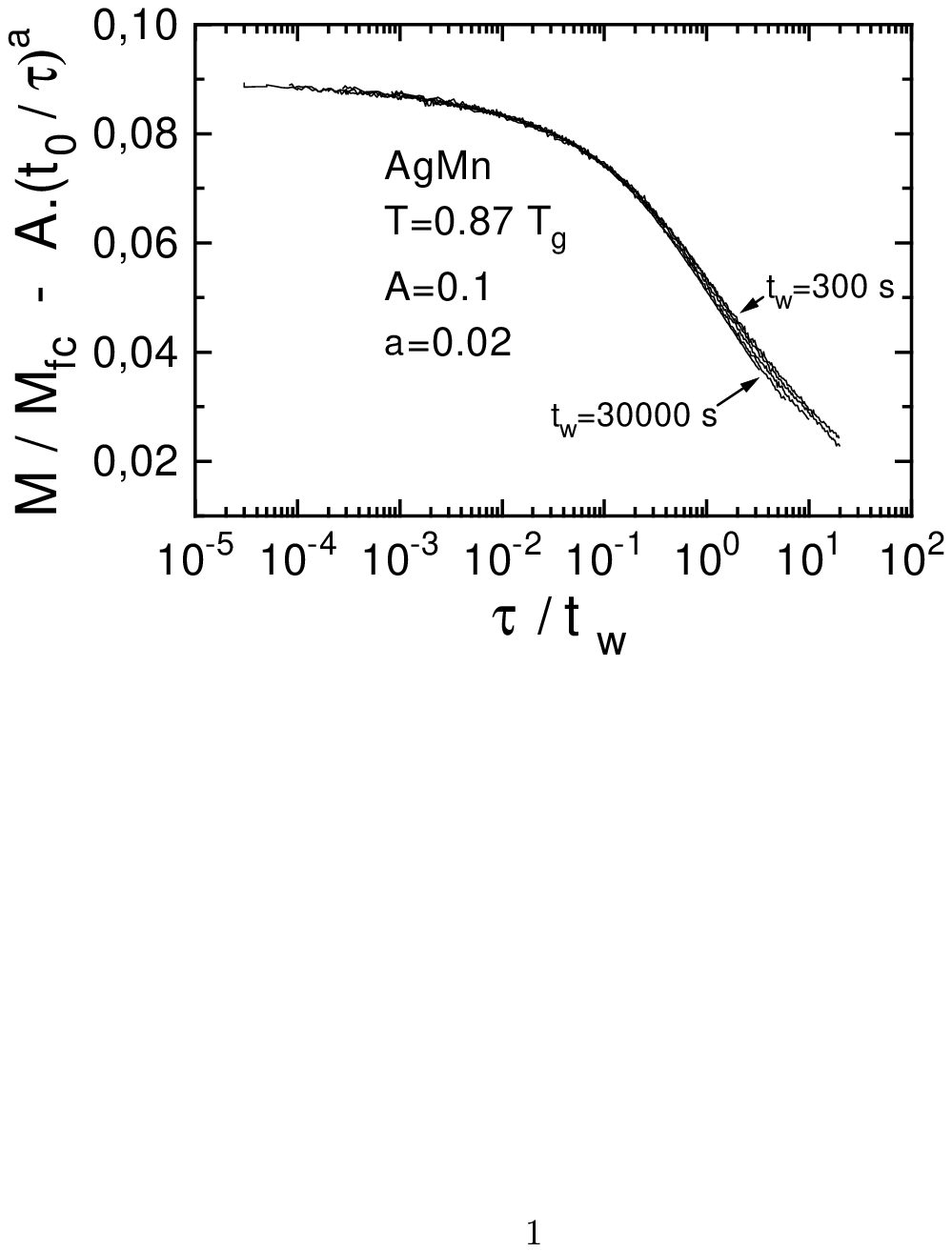}
}
\caption{The aging part of the
thermoremanent magnetization $M_{\sc ag}(t_w+\tau,t_w)$
(normalized by the zero field cooled value $M_{\rm fc}$)
vs. $\tau/t_w$ for $Ag {Mn}_{2.6}$ in a $\log_{10}$ scale
(from Ref. [14]).
The sample was cooled in a $0.1 \ Oe$ field from above the glass transition
$T_g=10.4K$
to a subcritical temperature $T=9K$. It waited  for
$t_w=300,1000,3000,10000,30000$ sec. under the
field that was suddenly switched off at $t_w$.
The decaying magnetization was recorded during all subsequent times $\tau+t_w$.
}
\label{revfigmexp}
\end{figure}

Another set of experiments which basically carry the same information is those
of the so-called `Thermo-Remanent Magnetisation' (TRM) relaxation
\cite{saclayrev1,saclayrev2}.
The system is cooled under a small magnetic field $H$, which is left
from $t=0$ (the time of the quench) to $t=t_w$, and then suddenly switched off.
The subsequent
relaxation of the magnetisation $M$ can be decomposed as
\cite{Bontemps,saclayrev2}
\be
M(t_w+\tau,t_w) = M_{\sc st}(\tau) + M_{\sc ag}(t_w+\tau,t_w)
\qquad\qquad
M_{\sc st}(\tau) \equiv \lim_{t_w \to \infty} M(t_w+\tau,t_w) \label{M}
\; ,
\label{mexp}
\ee
where again, there is a `fast' \footnote{It is fast in the sense that
experimentally, the major
part of $M_{\sc st}(\tau)$ has decayed to zero after the first second. However,
$M_{\sc st}(\tau)$ only decays as $(t_0/\tau)^a$ with $a$ small: see after Eq.
(\ref{chi}) above.
}
stationary contribution $M_{\sc st}(\tau)$
which is independent of $t_w$, and an aging part $M_{\sc ag}(t_w+\tau,t_w)$
which to a
good approximation (see Fig. \ref{revfigmexp} and Section
\ref{sectionmeanfield}) is a function
of the ratio $\tau/(\tau+t_w)$.
This again suggests that the effective relaxation time of the system is of the
order of its age. Actually, within
the linear response approximation, $\chi(\omega,t_w)$ and $M(t_w+\tau,t_w)$ are
essentially Fourier transform of
each other. More precisely, introducing the response function $R(t,t')$, one
has
\be
M(t_w+\tau,t_w) = H \int_0^{t_w} dt' \ R(t_w+\tau,t')
\; ,
\qquad\qquad
\chi(\omega,t_w)= \int_0^{t_w} dt' \ R(t_w,t') e^{i\omega(t'-t_w)}
\; .
\ee
TRM relaxation in spin-glasses \cite{saclayrev1}
and stress relaxation \cite{St,ElecGlasses}, electric polarisation
\cite{Gilchrist,Levelut}
or specific heat
\cite{Biljakovic} relaxation in many very different glassy materials show --
rather remarkably --
similar features, with a fast initial drop at
small times $\tau$, followed by a slow decrease of the signal on time scales of
  the order of the waiting time $t_w$. The same picture also pertains
-- on much smaller time scales -- to
numerical simulations of the response function of the three- \cite{Anma,Rieger}
and four-
dimensional \cite{Pariru} Edwards-Anderson and in mean-field
\cite{Cukuri}$^-$\cite{sknum}
spin-glass models.

Since the response function depends on some aspects of the thermal history, it
is interesting to consider more complicated experimental protocoles
such as different cooling rates from high temperatures to
$T_1$,  temperature cycles \cite{tempcycles1,tempcycles2}
between two temperatures $T_1$ and $T_2$, or even
field-cycling \cite{fieldcycles}. The detailed discussion of these
situations is
beyond the scope of the present paper. However, it is interesting to notice
that while the a.c. susceptibility $\chi(\omega,t_w)$ depends extremely weakly
on the cooling rate in spin-glasses, there are experimental systems (e.g.
dipolar glasses \cite{Levelut}) for which this dependence is large. We shall
come back to this point in Section \ref{RFIM}.

Finally, let us mention that aging can also be seen in {\it correlation
functions} (rather than response
functions). In equilibrium, these two quantities are related by the well-known
Fluctuation-Dissipation Theorem ({\sc fdt}), which is, as we shall discuss
in Section \ref{sectionfdt}, not necessarily valid in out of equilibrium
situations: in general,
correlation and response do not contain the same information.

 From an
experimental point of view, correlations are obtained from  time dependent
noise spectrum measurements \cite{Miguel,Weissmann} $S(\omega,t_w)$,
 which are rather more difficult than response
measurements {\footnote{This requires to perform many independent quenches
where the magnetic noise is recorded for different ages $t_w$ and then averaged
over the different quenches.}}. From a numerical point of view, however, it is
very easy to compute the time dependent correlation function $C(t_w+\tau,t_w)$,
which for a spin system is defined as
$C(t_w+\tau,t_w)=\frac{1}{N} \sum_{i}
S_i(t_w+\tau) S_i(t_w)$. The behaviour of $C(t_w+\tau,t_w)$, obtained on
relatively
short time scales from simulations for finite dimensional
\cite{Anma}$^-$\cite{Pariru}
and mean-field \cite{Cukuri}$^-$\cite{sknum}
spin-glass models, reveals aging below the spin-glass transition (for reviews,
see \cite{Riegerrev,Maparu}), again
qualitatively described by a shape similar to Eq. (\ref{M}), with $M$'s
replaced by $C$'s.

\subsection{Think in the two times plane : time sectors.}
\label{Timesectors}

Out of equilibrium situations have long been considered as untrustworthy. What
the experiments tell us (and what the theories below will confirm) is that
provided one abandons the idea that the correlation or response functions
should be `time translational invariant' ({\sc tti}), one can make sense of the
experimental data by explicitly keeping the dependence on the two times:
$C(t_w+\tau,t_w) \neq C(\tau)$. Equivalently, the Fourier transform of these
quantitites will not be functions of the frequency $\omega$ only, but of both
$\omega$ and the time
since the quench $t_w$.

More precisely, for any physical system there are a priori two other time
scales, one of them is microscopic (and will be noted $t_o$) and determines
for example, the single spin-flip time. The other time scale is the
equilibration time $t_{\sc erg}$, which, for a finite size system, will
always be finite (albeit often astronomical). The regime in which one expects
to observe some `universal' features (independent, for example, of the details
of
microscopic dynamics) is the regime where:
\be
t_o \ll t_w+\tau \ll t_{\sc erg} \ \ \mbox{and} \ \ t_o \ll t_w \ll t_{\sc erg}
\; .
\ee
This does not require any particular relation between $\tau$ and $t_w$. For
$t_{\sc erg}=\infty$, one may in general expect (and one indeed finds in some
models) that the functional form of -- say -- the correlation depends on the
way $t_w+\tau$ and $t_w$ are taken to infinity.

The simplest example to see this is the case of ferromagnetic domain growth.
After
a quench from high temperatures to a non-zero temperature below the Curie
point,
a pattern of domains of positive and negative magnetizations starts coarsening.
The typical
domain size  $\xi(t_w)$ at time $t_w$ diverges as a power law
 (or possibly more  slowly in the presence of impurities -- see Section
\ref{RFIM}).
 The ergodic time $t_{\sc erg}$ is then the
time at which the size of the domains is that of the sample. We shall consider
the thermodynamic limit in which $t_{\sc erg} = \infty$. Within an approximate
(large $n$)
theory of coarsening, the correlation function for large times is indeed found
to be
\cite{Br}
\be
C(t_w+\tau,t_w) = C_{\sc st}(\tau) + C_{\sc ag}\left(\frac{\xi(t_w)}
{\xi(t_w+\tau)} \right)
\; .
\label{C}
\ee

The first term describes the fast relaxation of the spins within each domain,
and has the same form as it would have in equilibrium, when there is only one
infinite
domain. Its limit\footnote{Note the crucial ordering of the two limits.}:
\beq
q_{\sc ea}  \equiv \lim_{\tau \rightarrow \infty} \lim_{t_w \rightarrow \infty}
C(t_w+\tau,t_w)
\label{qEAdef}
\eeq
is an example for this simple model of the Edwards-Anderson parameter,  a
quantity that
 plays an important role
in glassy dynamics (in this case it is simply the magnetization squared).
 The second term of Eq. (\ref{C}) describes the relaxation  of the system due
to the motion of
domain walls and  it manifestly depends on the waiting time.

As $t_w$ goes to infinity, one will be probing two distinct regimes,
depending on whether one takes $t_w \rightarrow \infty$ with $\tau$ finite
(`stationary' regime) or $\tau, t_w \rightarrow \infty$ with
\linebreak $\xi(t_w)/\xi(t_w+\tau)< 1 $ (`aging'
or coarsening regime).
In terms of the correlation function, the stationary and aging regimes are
simply
defined as the regimes of (large) times in which $C(t_w+\tau,t_w)>q_{\sc ea}$
and
$C(t_w+\tau,t_w)<q_{\sc ea}$, respectively.

These considerations can be translated in Fourier space as follows. Defining
\be
\hat C(\omega,t_w)=\int_0^{t_w} d\tau \ C(t_w,t_w-\tau) \ e^{i\omega \tau}
\; ,
\ee
one finds that $\hat C(\omega,t_w)$ does not only depend
upon $\omega$ (as it would in equilibrium situations) but also upon
$t_w$. For example, for the correlation
of the form (\ref{C}), one obtains, in the limit $\omega t_w \gg 1$:
\beq
\hat C(\omega,t_w) =  \hat{C}_{\sc st}(\omega) + \frac{1}{\omega} \; {\cal C}
\left(\frac{\xi(t_w) \, \omega}{\xi'(t_w)}\right)
\eeq
where ${\cal C}$ is a certain function related to $C_{\sc ag}$. In the simplest
case
in which $\xi$ grows as a power law, one finds
$\xi(t_w)\omega/\xi'(t_w)=\omega t_w$.

More generally, one can  envisage the possibility that different
physical mechanisms act on different large-time sectors,  defined as
\beq
\frac{\tau}{t_0}=O(1) \; ,  \;\;\; \;\;\;  \;\;\; \;\;\;
\frac{h_1(t_w)}{h_1(t_w+\tau)}=O(1)
\; ,  \;\;\; \;\;\;   \;\;\; \;\;\;
 \frac{h_2(t_w)}{h_2(t_w+\tau)}=O(1)
\; ,  \;\;\;\;\;\; {\mbox{etc.}}
\label{scaling}
\eeq
where the different functions $h_i$ (no longer necessarily related to domain
sizes)
are monotonously increasing functions which grow differently,
in such a way that:
\beq
0<\frac{h_i(t_w)}{h_i(t_w+\tau)}<1 \;\;\;\; \Rightarrow \;\;\;\;
 \frac{h_j(t_w)}{h_j(t_w+\tau)}=  1 \;\;\;\; {\mbox{for}} \;\; i <  j
\label{domi}
\eeq
Notice that these time sectors correspond to
asymptotically distinct relative `epochs' in the sense that if $t_1,t_2$
belong to the domain defined by $h_i$ and $t_2,t_3$ to the one defined by $h_j$
with $j>i$,
then $t_1,t_3$ also belong to the sector by $h_j$.

The correlation function can be, for example, the sum of terms
of the type (\ref{C}), each with a different scaling function $h_i$
replacing $\xi$. Because the different scalings vary in time-sectors that do
not overlap, such a function cannot be reexpressed in terms of a simpler
scaling form
valid for all large times.

A simple example  for the $h_i(t)$ is
\be
h_i(t)=
\exp \((\frac{t^{1-\mu_i}}{(1-\mu_i)t_o^{1-\mu_i}}\))
\ee
with $0 \leq \mu_i \leq 1$. In this case the $i^{th}$ scaling form  corresponds
 to $\tau \sim t_o^{1-\mu_i}
t_w^{\mu_i}$. In particular, $\mu=0$ yields the time-translational invariant
form
and $\mu=1$ the $\tau/t_w$ scaling variable which is independent of the
microscopic time $t_0$.

The main message of this Section is  that once one abandons time-translational
invariance,
 as one should in systems that never equilibrate, the two-time correlation (or
response)
 function in the long-times limit may have a
rich structure  including multiple-scaling forms like (\ref{scaling}).
 It will turn out that a rather general classification of the asymptotic
behaviour in the
 two-time plane  can be made using only  the monotonicity property
of the correlation and simple group theory (see Section \ref{solution}).

 The simple example of coarsening also illustrates the fact that in order to
decide
whether a system can be considered to be in equilibrium,
 {\em one-time} quantities (such as the energy, magnetization, etc) can
be misleading. Indeed, at any finite time the excess energy density  of a
coarsening
ferromagnet is proportional to the total domain surface divided by the volume,
a
quantity which soon  becomes very small. If one were  to judge the degree of
equilibration by only measuring the excess energy density, one would wrongly
conclude that the system equilibrates rather rapidly. On the other hand, two
time quantities (such as the correlation function) reveal very clearly that the
system is still out of equilibrium even at long times.

\subsection{Fluctuation-Dissipation Relations}
\label{sectionfdt}

As was mentioned before, in a system at equilibrium, the response to an
external
magnetic field and
the autocorrelation
functions are related through the fluctuation-dissipation theorem ({\sc fdt}).
This is true in general for the response to a field $h$ conjugate to any
observable $O$
and the corresponding autocorrelation $C_{\sc o}(t_w+\tau,t_w)
\equiv \langle O(t_w+\tau)O(t_w)\rangle$.
In equilibrium:
\beq
R_{\sc o}(t_w+\tau,t_w) \equiv \left. 
\frac{\delta \langle O(t_w+\tau)\rangle}{\delta
h(t_w)}
\right|_{h=0}
 =R_{{\sc o}, {\sc eq}}(\tau)=
-\frac{1}{T} \frac{\partial C_{{\sc o},{\sc eq}}(\tau)}{\partial \tau}
\label{fdt1}
\eeq
or, introducing the integrated response $\tilde \chi_{\sc
o}(t_w+\tau,t_w)=\int_{t_w}^{t_w+\tau}
R_{\sc o}(t_w+\tau,t') dt'$:
\beq
\tilde \chi_{\sc o}(t_w+\tau,t_w) = \tilde \chi_{{\sc O},{\sc eq}}(\tau) =
 \frac{1}{T} (C_{{\sc O},{\sc eq}}(0)-C_{{\sc O},{\sc eq}}(
\tau))
\; .
\label{fdt2}
\eeq
If $O$ is the energy, one obtains the relation between energy fluctuations and
specific heat, if $O$
is the magnetization, one finds a relation between the
(time dependent) field induced magnetisation and the magnetic noise
correlations, etc.

In order to study the relation between `fluctuations' and `dissipation'
in out-of-equilibrium systems, one has to think in terms of two-time
correlation and response functions.
Let us then consider a given $t_w$ and make a parametric plot of $\tilde
\chi_{\sc o}(\tau+t_w,t_w)$ {\em vs.}
$C_{\sc o}(\tau+t_w,t_w)$ when $\tau$ varies.
One then takes a larger $t_w$ and repeats the plot, and so on.
If the system equilibrates after a finite
time $t_{\sc erg}$, one obtains, when $t_w \gg t_{\sc erg}$, a limiting
$\tilde \chi_{\sc o}$ {\em vs.} $C_{\sc o}$ curve which is
a straight line with slope $-1/T$: this is the {\sc fdt}.

\begin{figure}
\centerline{\epsfxsize=9cm
\epsffile{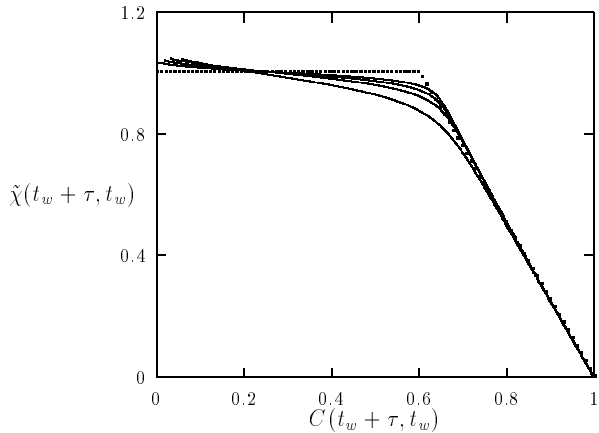}
}
\caption{
The susceptibility $\tilde\chi(t_w+\tau,t_w) \equiv \int_{t_w}^{t_w+\tau} ds
\, R(t_w+\tau,s)$ vs the correlation $C(t_w+\tau,t_w)$
for domain growth in a $n$-component vector ferromagnet at $T < T_g$, obtained
from
an analytical treatment of the large $n$ limit. (From Ref.[48].)
} 
\label{p2chitfinite}
\end{figure}

Consider instead what happens in the example of domain growth in an infinite
size system. Within a large $n$
treatment of the problem \cite{Cude1},
one obtains for the $\chi$ {\em vs.} $C$ curves a family of curves shown
in Fig. \ref{p2chitfinite}. For large  $t_w$,
the curves approach a {\it broken line}: one with
slope $-1/T$ for values of the correlation larger than $q_{\sc ea}$ ({\em i.e.}
for times in the stationary regime defined in the previous Section),
and one with zero slope for values of the correlation smaller than $q_{\sc ea}$
(the aging regime). This
simple example illustrates that (despite
 the fact that the dynamics is very slow when $t_w \to \infty$) the system
cannot
be thought of as in a `quasi-equilibrium' state, for which  concepts from
equilibrium
are more or less valid: there is always a regime in which {\sc fdt} is strongly
violated.
(It would be interesting to confirm these results within more realistic
models of coarsening, and also, obviously,
experimentally and numerically.) We also see that the
Edwards-Anderson parameter  plays a
role for the response functions.

\begin{figure}
\centerline{\epsfxsize=9cm
\epsffile{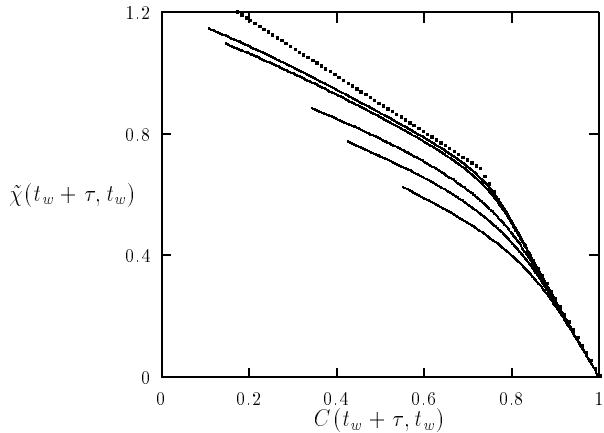}
}
\caption{
The susceptibility $\tilde \chi$ vs the correlation $C$ for the `discontinuous'
mean-field spin glass
model of Section \ref{sectionmeanfield} ($p=3$)
at $T< T_g$. In this case the {\sc fdt} violation for $C < q_{\sc ea}\sim 0.76$
is given by $-X/T$,
with $ X = (1-q_{\sc EA})/q_{EA}\sim 0.76$. More complicated situations are described in
Section \ref{sectionmeanfield}.
(From Ref.[48].)
}
\label{p3chitfinite}
\end{figure}

Similar results are found analytically in the mean-field spin-glass systems
which we shall
review below
(see Fig. \ref{p3chitfinite} and Section \ref{sectionmeanfield}).
The main difference is that the aging part of the curve (i.e. $C<q_{\sc ea}$)
has
a non-zero slope.\cite{Cuku1}$^-$\cite{Cule}
This is quite important, since it means that the integral
of the response function over the aging regime gives
a {\it non-zero} contribution.

In the more realistic $3D$ Edwards-Anderson model, the form of the  $\tilde
\chi$ {\em vs.} $C$ curves which are
obtained numerically  (at least for the computer times
accessible at present) are
actually remarkably similar \cite{Frrie}
 to the mean-field prediction for the corresponding mean-field model
\cite{Cuku2}.
The same is true of some very recent numerical simulations
of binary soft-sphere mixtures \cite{binary}.

The violation of {\sc fdt} can be parametrized by introducing a violation
factor
$X_{\sc o}(t,t')$ defined as
\beq
R_{\sc o}(t,t') \equiv \frac{X_{\sc o}(t,t')}{T} \frac{ \partial C_{\sc
o}(t,t')}{\partial t'}\label{Gfdt}
\; .
\eeq
Note  that we are differentiating with respect to the {\em smallest} time $t'$.
In analytic studies of mean-field systems,  one can furthermore show
 that for large times $X_{\sc o}$ depends on $t,t'$ only through the value of
the correlation function:
$X_{\sc o}(t,t')=X[C_{\sc o}(t,t')]$. In particular, when $C_{\sc o} > q_{\sc
EA,o}$, $X_{\sc o}=1$,
 and the {\sc fdt} is recovered.

It turns out \cite{Cukupe} that the `effective temperature'
\beq
T_{\sc o}^{\sc eff}(t_w+\tau,t_w) \equiv  \frac{T}{X_{\sc o}(t_w+\tau, t_w)}
\label{Teff}
\eeq
is precisely the temperature which would be read on a thermometer with response
time $\tau$ (or frequency $\omega \sim 1/\tau$) when connected to the
observable $O$ at time $t_w$.
A `fast' thermometer of response time $\tau \ll t_w$ will then probe the
stationary
regime for which $X_o=1$  and thus measure
the heat-bath temperature. This is the reason why glasses, although still out
of equilibrium after
many hours, feel as cold as the room they are in.


\subsection{Edwards-Anderson parameter, weak-ergodicity breaking and
clonation}
\label{WEB}

The situation we have described in the last two subsections
is one in which  equilibrium is not achieved, in the sense that configurations
are
not visited with a probability given by the Gibbs-Boltzmann weight.
However, this is not the main point: none of the above results can be explained
within a  {\em strong ergodicity breaking} scenario where the system
falls into a very long-lived metastable state, and achieves fast
equilibration within such restricted sector of phase-space.  A mixture of
hydrogen and oxygen, or a crystal
of diamond
are systems which are metastable, but for all dynamical purposes in
equilibrium.
As Feynman puts it: `Equilibrium is when all fast things have happened, and
slow things not yet'.
The out of equilibrium situations which are of interest to us are those where,
in a sense, `things keep happening on all time scales'.

More precisely, in  a spin-glass (or a coarsening problem) in the low
temperature phase the spins do,
 on average,
 remember for some time their  orientation, which leads to a non-zero
 Edwards-Anderson parameter $q_{\sc ea}= \lim_{\tau \to \infty} \lim_{t_w \to
\infty} C(t_w+\tau,t_w) $.
However, if
the waiting time is {\it finite}, the system is able
to escape arbitrarily far from the configuration it had reached at $t_w$,
leading to
\be
\lim_{t_w \to \infty} \lim_{\tau \to \infty} C(t_w+\tau,t_w) = 0
\ee
even in the low temperature phase where $q_{\sc ea} > 0$. This situation was
called `weak-ergodicity breaking' in Refs.[\cite{Bou92,Cuku1,Cuku2}].

It is important to note that the fact that a system undergoes weak (as opposed
to strong) ergodicity breaking for
infinite times does not mean that stable states do not exist in phase-space.
The dynamical behaviour depends on the choice of initial conditions.
The example of ferromagnetic coarsening is again
eloquent in this respect. There exist two true equilibrium states, and the
relaxation from
an initial condition close to one of them (e.g. all spins up) is fast and shows
no aging.
But in the relaxation from a random initial condition (e.g. a typical
configuration at
a temperature above $T_c$),  the system remains forever (in the
thermodynamic limit)  undecided
as to which state it will go to.

Another interesting question is the following: suppose that at time $t_w$ one
`duplicates' the spin system and evolves subsequently the two copies using two
independent thermal baths. (This process was called `clonation' in
Ref.[\cite{Ba}].) Will the two copies stay `close together' or conversely
will they
evolve independently, forgetting their common breed? This is measured by the
`overlap' function $Q(t_w+\tau,t_w+\tau)$ defined as \cite{damage}
\be
Q(t_w+\tau,t_w+\tau)= \frac{1}{N} \sum_{i=1}^N S^{(1)}_i(t_w+\tau)
S^{(2)}_i(t_w+\tau)
\qquad S^{(1)}_i(t_w) \equiv S^{(2)}_i(t_w) \ ,
\ee
where the superscripts $^{(1,2)}$ refer to the two copies. We shall see below
that the limit
\be
\lim_{t_w \to \infty} \lim_{\tau \to \infty} Q(t_w+\tau,t_w+\tau) = Q_\infty
\label{Qinfty}
\ee
can be zero or non zero, even if $q_{\sc ea}$ is non zero. This may serve to
distinguish different types of aging dynamics \cite{BBM,Cude1,Ba,sknum}.

 If the limit is taken in
 reversed order (i.e., $t_w \to \infty$ first), the overlap function contains
 the same information as the correlation function. More precisely \cite{BBM}:
 \be
 \lim_{t_w \to \infty} Q(t_w+\tau,t_w+\tau) \equiv C_{\sc st}(2\tau)
\; .
 \ee

\section{Phenomenological models of aging}

In order to account for the above experimental results, one has as usual the
choice between some physically motivated, but phenomenological, pictures or
some rather more precise microscopic models, in a limit in
which they are analytically tractable. Both approaches are actually
complementary, and
shed light on each others' limitations. We shall
 start by reviewing
the phenomenological pictures of aging, based either on domain growth
arguments, or on models of random walks in phase space.

\subsection{Coarsening in non disordered systems}

 As we  noted in the previous Section, the simplest model where aging occurs is
the
ferromagnetic Ising model
suddenly quenched below its Curie point temperature. The initial configuration
is
random, and it orders progressively through domain growth. Depending on the
class of microscopic dynamics \cite{Br}, the typical size $\xi$ of the domains
grows
as $t^{1/2}$  (`non-conserved' case) or $t^{1/3}$ (`conserved' case).
The `age' of the system is thus directly encoded in the spatial correlation
functions. The two-time correlation function $C(t_w+\tau,t_w)=\frac{1}{N}
\sum_{i}
S_i(t_w+\tau) S_i(t_w)$ can be calculated exactly in some cases (e.g. the Ising
model in one dimension, or the large-$n$ `spherical' model), or using some
approximations (for a review see Ref.[\cite{Br}]). One finds an expression as
Eq.(\ref{C})
\be
C(t_w+\tau,t_w) = C_{\sc st}(\tau) + C_{\sc ag}
\left(\frac{\xi(t_w)}{\xi(t_w+\tau)}\right)
\ee
where $C_{\sc ag}$ decays as a (non trivial) power-law for large arguments.
$C_{\sc ag}(1+u)$ is in general also singular around $u=0$, its behaviour is
characterised by an
exponent \cite{Cule} $b$ and, in
the `non-conserved case', one can argue that \cite{Sire,DDJP},
$C_{\sc ag}(1)-C_{\sc ag}(1+u) \propto \sqrt{u}$.
Correspondingly, the aging part of the a.c. susceptibility
decays as in Eq. (\ref{chi}), with $b=1/2$.

The overlap function $Q(t_w+\tau,t_w+\tau)$ can also be estimated in this
simple
coarsening situation. Within standard approximate treatments of coarsening
\cite{Cude1,BBM}, one finds that the quantity $Q_\infty$
 defined in Eq.(\ref{Qinfty}) is non zero, meaning that the two copies follow
each others'
footsteps (within  finite times) in their evolution towards equilibrium.
Note also that within the large n approximation, the {\sc fdt} violating factor
$X$ goes to zero at infinite times, as $\xi^{-1}$.
It would be interesting to know whether this is a more general property of
coarsening dynamics.

It is interesting to remark that in the presence of a small external magnetic
field $H$, aging is `interrupted' after a finite time $t_{\sc erg}(H)$,
since one of the two phases is favoured by the field. For example, in the case
of the spherical model, one finds \cite{Cude2}
$t_{\sc erg}(H)\propto H^{-2}$. This
behaviour is expected on general grounds: it corresponds to the time beyond
which the curvature induced driving force is superseded
by the field induced driving force.

\subsection{Coarsening in disordered systems I: Random ferromagnet or random
field}
\label{RFIM}

Let us now consider the case of a disordered ferromagnet in dimension larger
than 2, where random local
magnetic fields or random local couplings are present. If the disorder is
sufficiently weak (for example, if a small fraction of the ferromagnetic
bonds are removed), the ground state of the system still has long-range
ferromagnetic order, and the description in terms of the growth of ordered
domains is valid. However, due to the presence of disorder, the domain
{\it walls} will tend to be pinned by local inhomogeneities. The problem
of domain walls in disordered environments has been the focus of intense study
 in the recent years \cite{domainwalls}. In many aspects, this problem is close
to the spin-glass
problem, with a large number of metastable states (although of course the
`spin-glass' nature of the problem only concerns the small fraction of spins
which belong to the domain walls). The dynamics of a given section of a domain
wall proceeds by thermally activated hops between different favourable
configurations. Ordered domains thus grow on average, but at a much reduced
rate compared to the pure ferromagnet described above.

A generally accepted description is as follows
\cite{RFIMscaling}: the
typical pinning energy scale of a domain wall of linear size $R$ grows as
$\Upsilon R^\theta$, where $\theta$ is an exponent which depends on the problem
(random field/random bond) and the dimension of space. A simple scaling
argument{\footnote{This argument assumes that barrier heights between
metastable states behave (as a function of $R$) similarly to the pinning energy
of each state \cite{DrosselKardar,VinokurIoffe}}} then suggests that the time
needed
for a domain to reach a certain size $R$ is given by
\be
\tau(R) \propto t_o \exp \left(\frac{\Upsilon R^\theta}{kT} \right)
\; .
\ee
After a certain time $t$, the typical size of the domains is thus expected to
be given by
\be
\xi(t) \propto \left(\frac{kT \log\left(
\frac{t}{t_o} \right) }{\Upsilon}\right)^{\frac{1}{\theta}} \; ,
\ee
provided the corresponding
pinning energy $\Upsilon \xi(t)^\theta$ is large compared to $kT$. (In the
other limit, the pinning energy is negligible, and one recovers the growth law
which we discussed in the previous paragraph.) This logarithmic growth of
domain sizes has been rather carefully checked numerically \cite{RFIMnum}  in
$D=3$ ($T \neq 0$).

Apart from the fact that domains
grow very slowly, one  expects that the picture prevailing in the pure
ferromagnetic case is not drastically modified. In other words, the correlation
function should still be given by Eq. (\ref{C}), but with a logarithmically
growing $\xi(t)$. This was confirmed numerically on the Ising
model with random fields, for $D=1$ \cite{DDJP} and $D=3$ \cite{Levelut-us}. In
$D=1$, the motion of a `domain
wall' (a point) is given by Sinai's diffusion law, i.e. $\xi(t) \propto
\log^2(t)$ (at least for times smaller
than a certain temperature dependent $t_{\sc erg}$, which diverges
when $T \to 0$). The aging part
of the two-time correlation function can indeed be satisfactorily rescaled when
plotted versus $\xi(t_w)/\xi(t_w+\tau)$.

It is important to notice that the above scenario, where the system tries to
reach a well defined (temperature independent) state, but is slowed down due
to pinning by impurities, leads to large cooling rate effects. This is because
the crossover energy $\Upsilon \xi(t)^\theta \simeq kT$ will be reached at
later times
if the cooling rate is slower \cite{Levelut-us}. The size of the domains when
the working temperature $T_1$ is reached will thus be larger,
 the smaller the cooling rate, and this will affect many physical observables,
 such as the energy.

Let us finally point out that disorder is actually not necessary to obtain
logarithmic in time (rather than power-law) growth of the domain size. This was
first shown in
Ref.[\cite{Sethna}] for a pure Ising model with next-nearest neighbour
couplings.
Actually, if the domain walls are below their roughening temperature, the
dynamics
proceeds via the nucleation of terraces\cite{NozieresGallet}, and also leads
to logarithmic domain
growth.

\subsection{Coarsening in disordered systems II: Spin glasses and droplets}
\label{DROPLETS}

It is not obvious whether the above coarsening description also applies
to spin glasses, because of the non-conventional nature of their order
parameter.
However, Fisher and Huse \cite{FH} have argued that for
spin-glasses in
finite dimension and for any given temperature below the spin-glass transition,
there
are only two `pure states' (spin-reversed from each other) which have to be
considered, which can conventionally be called `up' and `down', very much like
in the
Mattis model \cite{mattis}. This assumption gives some physical content to a
scaling description of the spin glass phase, first advocated in
\cite{McM,BrayMoore}. It allowed
several authors \cite{FH,KH,Thill} to develop a rather complete
 phenomenological picture of spin-glasses in low dimensions, where the
spin-glass is considered
as a `disguised ferromagnet' (with however the important
difference
that the two pure states are not stable when the temperature is changed -- see
below).  This is at variance with the mean-field picture emerging from Parisi's
solution \cite{Pa} of the Sherrington-Kirkpatrick model \cite{Shki}, where many
(non trivially
related) pure
states coexist \cite{Mepavi,MPSTV}

If this `two-state' picture is retained, the dynamics of the system can again
be
described in terms of growing and coalescing compact domains \cite{KH},
which have also been
called `droplets' \cite{FH} in this context.
The presence of disorder presumably pins the
domain
walls, leading again to a logarithmic growth of the droplets, and thus to a
two-time
correlation very similar to the random-bond or random field case described
above. In
particular, the aging part of the correlation function should be a function of
{\footnote{Fisher and Huse actually postulated
that the
scaling variable would rather read
$\xi(\tau)/\xi(t_w)$.}}
$\xi(t_w)/\xi(t_w+\tau)$.

One striking property of  experimental
spin-glasses, however, is the very weak dependence of
its
physical properties on the cooling rate \cite{private}. For example, the
asymptotic value of
the
a.c. susceptibility, $\chi(\omega,t_w \to \infty)$, is nearly independent of
the
cooling rate. This is at first sight surprising in a scenario of activated
domain
growth, as we emphasized above. But if one argues that the spin-glass phase is
`chaotic' \cite{BrayMoore,FH,KH}, i.e. that the two pure states towards which
the system evolves are extremely
fragile to temperature changes, it becomes obvious that the dynamics of the
system
at temperatures greater than $T_1$ is useless to bring the system closer to its
equilibrium at the working temperature $T_1$. In a first approximation, the
configuration reached by the system at temperature $T_1 + \delta T$ is as
remote
from the `true' equilibrium state at temperature $T_1$ than a high temperature
configuration. Hence the cooling rate may indeed have a negligible effect.

How does the two-state picture compare with aging experiments or numerical
simulations, in the case of a $3-d$ Ising spin-glass? The detailed discussion
of
this
point is beyond the scope of the present paper and the conclusions still
controversial \cite{fieldcycles,Nosvrev,NewmanStein2,Maparu},
in particular the existence of a transition in a magnetic field
\cite{ATline1}$^-$\cite{Myd}.
One should however mention the following results:

$--$
Both experimentally and numerically, the aging part of the
correlation and
response functions follow a scaling which is systematically closer to
$\tau/t_w$
than the
expected $\xi(t_w)/\xi(t_w+\tau)$ (or even $\xi(\tau)/\xi(t_w)$)
if the
barrier heights scaled as $R^\theta$, e.g.
$\xi(t_w)=\log(t_w/t_o)^{1/\theta}$.
The point is that such a scaling would place the curves of Fig. 2 in the
reversed order:
the young ones would be {\it below} the old ones in a $\tau/t_w$ plot.

However, as suggested by Rieger \cite{Riegerrev}, the two state picture with
barrier heights scaling as
$\Upsilon \log R$ (i.e. $\theta \to 0$) would lead to an algebraic domain
growth law (as postulated
by Koper and Hilhorst \cite{KH}): $\xi(t) \propto t^\alpha$ with
$\alpha=kT/\Upsilon$ \cite{Rieger,Maparu}, and thus in turn to a $\tau/t_w$
scaling.

$--$ A direct numerical indication of a growing length scale was searched
for in
\cite{Huse,Rieger}. A possibility is to study the following
correlation function
\be
G(\vec r,t) = \frac{1}{N} \sum_i
\langle
S_i^{(1)}(t) S_i^{(2)}(t) S_{i+\vec
r}^{(1)}(t)
S_{i+\vec r}^{(2)}(t)
\rangle
\; ,
\ee
where $S_i^{(1)}$ and $S_i^{(1)}$
which has the following intuitive meaning: knowing that the copies $1$ and $2$
are in
the same state at site $i$ (resp. opposite states), what is the probability
that they
are still in the same state (resp. still in opposite states) a distance $\vec
r$
apart ? Numerically, $G(\vec r,t)$ is seen (in $3$ dimensions) to be of the
form
\cite{Marinari}:
\be
G(\vec r,t) = \frac{1}{r^\zeta} g\left(\frac{r}{\xi(t)} \right) \qquad \xi(t)
\propto
t^\alpha
\ee
which indeed suggests the presence of a growing scale $\xi(t)$ in the dynamics.
The
conventional droplet picture predicts (apart from a logarithmic, rather
than power-law, growth of $\xi(t)$) that $\zeta=0$, since the equilibrium state
should be unique up to a global sign change, whereas
simulations by Marinari {\it et al.} \cite{Marinari,Maparu} suggest that
$\zeta >0$, just as in the equilibrium situation obtained
with replica field theory \cite{deDoko} in dimensions smaller
than 6.

$--$ The temperature cycling experiments \cite{saclayrev1} where the system
is cooled to temperature
$T_1 < T_g$, then to temperature $T_2 < T_1$, and finally back to temperature
$T_1$, show a very
striking conjunction of `rejuvenation' (when the temperature is decreased) {\it
and} memory
(when the system is heated back). The coexistence of these two effects
is rather awkward to interpret within the droplet picture \cite{saclayrev1,BD}.
In
the same spirit, Weissmann et al. \cite{Weissmann}
 have argued that the `second noise spectrum' of spin-glasses does not conform
with what could be
 expected from a simple `two-state' picture.

In summary: even if the `disguised ferromagnet' picture did
provide a correct
description of the equilibrium properties of low-dimensional spin-glasses, it
is
 not obvious that this description is sufficient to account for the {\it
out-of-equilibrium} properties.
One reason is that -- precisely because of the chaotic nature of the
equilibrium phases -- the system will
 not only nucleate domains of the nominal equilibrium state, but probably also
`phases' corresponding to
 nearby temperatures, which will thus contribute to the non-stationary part of
the  dynamics.

\subsection{Stranded in phase space: the `trap' model}

`Phase-space' models are another very useful class of phenomenological models
for the
dynamics of complex systems, and have been advocated by very many authors over
the years\cite{Phasespacemodels,Phasespacemodels2}.
The dynamics of the whole system is summarized in
the
motion of a single point evolving within a complicated energy landscape in
configuration space.
 From a general point of view, one expects this energy landscape to be made of
`valleys', or `traps' (within which all configurations are mutually accessible
in a short time)
separated by `barriers', which the system can only overcome by thermal
activation. A
coarse-grained representation of the problem can thus be given in terms of
states
$\alpha,\beta,\gamma...$ between which the system wanders. The dynamics of the
system
is thus described by a master equation for the probability to find the system
in
the state $\alpha$:
\be
\frac{\partial P_\alpha}{\partial t} = - \sum_{\beta} W_{\alpha \to \beta}
P_\alpha
+ \sum_{\beta} W_{\beta \to \alpha} P_\beta
\; .
\ee
The choice of the hopping rates $W_{\alpha \to \beta}$ then encodes the
statistics of
the barrier heights and the geometry of the phase space.
It is rather arbitrary apart from the constraint of detailed balance.
 For example, one can
organize the `traps' on a hierarchical tree, and choose $W_{\alpha \to \beta}$
to
only depend on the distance between $\alpha$ and $\beta$ along the tree. This
has led
to several `ultrametric' diffusion models \cite{Phasespacemodels}, with many
interesting results, including,
in some cases, aging effects.\cite{Fevi}$^-$\cite{Melin}
\vskip 0.5cm
$\bullet$ The one-level tree.
\vskip 0.5cm

The simplest of these models,\cite{Bou92} for which the appearance of aging has a
particularly
clear interpretation, is when the hopping rate only depends on the starting
state:
$W_{\alpha \to \beta} =(N \tau_{\alpha})^{-1}$, where $N$ is the total
number of states. (The final state $\beta$ is thus independent of the initial
state;
the process starts anew at each jump.) This corresponds to the picture drawn on
the left of
Fig. \ref{figjptree1}. The
trapping times $\tau_{\alpha}$ are of the form $t_o
\exp(B_\alpha/(kT))$, where $B_\alpha$ is the energy barrier
`surrounding' state $\alpha$. Within this description, the {\it equilibrium}
measure $P_\alpha^{eq}$ (if it exists) is simply proportional to
$\tau_\alpha$. In order to reproduce the correct Bolzmann equilibrium, one
should thus identify $B_\alpha$ with the free-energy of the state{\footnote{The
relation
between barrier heights and energy depths is however
not obvious in general; for recent work on this subject, see Ref.
[\cite{DrosselKardar}]}
 $f_\alpha$.
Mean-field models of spin-glasses \cite{randfree} or replica treatment of
randomly pinned manifold suggest that the distribution of
the metastable states' free energies $f_\alpha$ is
exponential \cite{randfree,Mepavi,PHouches,Balents}
\be
\rho(f_\alpha) \propto \exp\left(-\frac{x |f_\alpha|}{kT} \right)\label{rhof}
\ee
with a certain parameter $x \leq 1$ in the glassy phase, $x =T/T_g$
in the Random Energy Model\cite{BREM,GREM}. The appearance of this exponential tail for `deep'
states can be understood on general grounds, and is related to the so-called
`extreme value statistics' (for a more precise discussion, see
Refs. [\cite{Vinokur,BMinprep}]).
The corresponding distribution of trapping times {\footnote{Generalisation to
other distributions of $f_\alpha$ has been considered in Ref.[\cite{Monthus}]}
is
then easily found to be
\be
\rho(\tau) d\tau = \rho(f) df \longrightarrow \rho(\tau) \propto_{\tau \gg
t_o} \frac{t_o^x}{\tau^{1+x}}
\; .
\ee
Let us now introduce the quantity $\Pi(t_w+\tau,t_w)$ defined as the
probability
that the system has not changed trap between time $t_w$ and time $t_w+\tau$.
This
quantity is found to be very different depending on whether $x$ is larger or
smaller than 1. In the former case, $\lim_{t_w \to \infty} \Pi(t_w+\tau,t_w)$
is
well defined, and found to be proportional to $(t_o/\tau)^{x-1}$. For
$x < 1$, however, one finds that $\Pi(t_w+\tau,t_w)$ `ages', and is given by:
\be
\Pi(t_w+\tau, t_w) = {\sin(\pi x)\over \pi}\int_{{\tau\over \tau + t_w}}^1
du(1-u)^{x-1} u^{-x} \qquad (t_w \gg t_o)\label{pi}
\; .
\ee
In physical terms, this means that after a waiting time $t_w$, the only states
which have an appreciable probability are those with a trapping time of the
order
of $t_w$ itself. This reflects the fact that the distribution of trapping times
$\rho(\tau)$ becomes so broad when $x < 1$ (the average trapping time becomes
infinite), that the sum of all the trapping events $\tau_1+\tau_2+ ...+ \tau_N$
is always dominated by its largest term, which is thus of the order of the
experimental time itself \cite{Bou92}. This dominance of a few very important
events is a characteristic feature of L\'evy statistics \cite{PhysRep}. On the
other hand,
when $x > 1$, the trapping times are all of order $t_o$.

\begin{figure}
\centerline{\epsfxsize=9cm
\epsffile{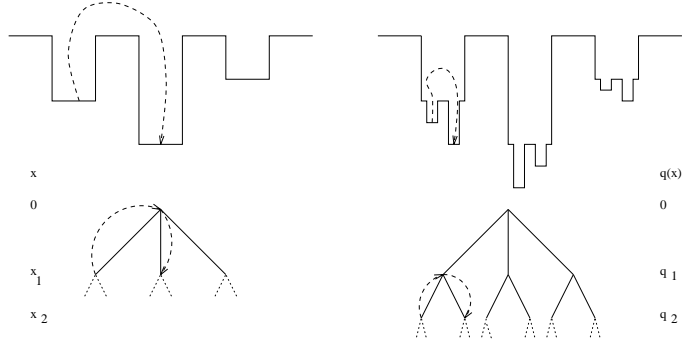}
}
\caption{Schematic phase-space landscape of a one level tree and of a
multi-level
tree.
}
\label{figjptree1}
\end{figure}

Let us now define the overlap $q_{\alpha \beta}$ between states as $q_{\alpha
\alpha} = q_{\sc ea}$ and $q_{\alpha \neq \beta}=q_0$ (more general choices
will
be discussed below). The self-overlap $q_{\sc ea}$ is smaller than $1$ in
general;
this reflects the fact that many microscopic configurations are mutually
accessible within times of order $t_o$; this contributes to the equilibrium
part of the
correlation function which the one-tree level cannot describe. The  spin-spin
correlation function,
averaged over the disorder \footnote{
i.e. over the distribution of $f_\alpha$. Note that in this model, the
correlation function is not
self-averaging, precisely for the same reason as for the static overlap
distribution $P(q)$ in mean-field
spin-glasses \cite{MPSTV}} is thus given by
\be
C_{\sc ag}(t_w+\tau,t_w) = q_{\sc ea} \Pi(t_w+\tau, t_w) + q_0 (1-\Pi(t_w+\tau,
t_w))
\; .
\ee
Note in particular that $\lim_{\tau \to \infty}$
$\lim_{t_w \to \infty} C_{\sc ag}(t_w+\tau,t_w)
$ $=$ $q_{\sc ea}$ when $x < 1$, but that \newline
 $\lim_{t \to \infty} \lim_{t_w \to \infty}
C_{\sc ag}(t_w+\tau,t_w) = q_0$ for $x > 1$. Within this model, $x=1$ thus
corresponds to a
true glass transition.

Equation (\ref{pi}) leads to the following asymptotic
behaviour for $C_{\sc ag}(t_w+\tau,t_w)$ \cite{BD}:
\begin{eqnarray}
C_{\sc ag}(t_w+\tau,t_w) &\simeq& q_{\sc ea} - \frac{\sin(\pi x) (q_{\sc
ea}-q_0)}{\pi (1-x)}
 \left({\tau \over t_w}\right)^{1-x}
\qquad
(\tau \ll t_w) \label{Cshort}
\\
&\simeq& q_0 + \frac{\sin(\pi x) (q_{\sc ea}-q_0)}{\pi x}\left({t_w\over \tau
}\right)^x
 \qquad \qquad
(\tau \gg t_w) \label{Clong}
\eea
Hence, both the `short time' $(t_o \ll \tau \ll t_w)$ and `long time' regimes
are described by power-laws, much like in the simple coarsening models
described in Sections \ref{RFIM} and \ref{DROPLETS}.

The above model can be endowed with magnetic properties by assigning to each
state a certain
magnetisation $m_\alpha$, and modifying the hopping rates in the presence of a
field to recover
the correct equilibrium weights \cite{BD}. One then finds that the generalised
form of the {\sc fdt} (Eq. (\ref{Gfdt})) holds, with:
\be
X(t_w+\tau,t_w) = 1 - \zeta + \zeta \frac{t_w}{t_w + \tau}
\ee
where $\zeta$ is a free parameter of the model, restricted to the interval
$[0,1]$. For $\zeta=0$, the thermoremanent magnetisation is simply proportional
to $C(t_w+\tau,t_w)$.

Experiments on the thermoremanent magnetisation or the a.c.
susceptibility show that both the `short time' and `long time'
regimes can be fitted by power-laws (see Fig. 7 in Ref.[\cite{saclayrev2}]). The
exponent $x$ which comes
out of these
two fits is however different: $x =1-b \simeq 0.6-0.9$ (depending on
temperature)
from the $\tau \ll t_w$ region, and $x \simeq 0.1-0.3$ from the $\tau \gg t_w$
region
\cite{BD,saclayrev2}.

One can also define the overlap function $Q(t_w+\tau,t_w+\tau)$ within such a
model,
and show that it is simply related to
$C$ through $Q(t_w+\tau,t_w+\tau)=C(t_w+2\tau,t_w)$. One thus finds, in
particular,
$Q_\infty=q_0$: two copies of the same
system decorrelate completely from each other \cite{BBM}. Note however that
this
would not be true if the `traps' were organized along a low dimensional `path'
in phase space.
\vskip 0.5cm

$\bullet$ The multi-level tree.
\vskip 0.5cm

An interesting generalisation of this `one-level' trap model is to consider a
hierarchical organisation of traps within traps, and to relate the overlap
$q_{\alpha \beta}$ between two states to their distance along the tree (see
the figure on the right of Fig. \ref{figjptree1}). The hopping rate $W_{\alpha
\to \beta}$ is still taken to be
independent of the final state for a given distance along the tree, or for a
given overlap $q=q_{\alpha,\beta}$
\be
W_{\alpha \to \beta} \propto \frac{1}{\tau_{\alpha,q}}
\; .
\ee
As reviewed in Refs [\cite{Mepavi,PHouches,MP}], the interpretation of the static
full replica symmetry breaking solution suggests that the distribution of
free-energies
of the states at a certain level of the tree is still exponential, but with a
parameter $x$ which now depends on the overlap between these states (and is the
inverse of the Parisi function $q(x)$)
\be
\rho_q(f) \propto \exp \left( -\frac{x(q)(\tilde f - f)}{kT} \right)
\; ,
\ee
where $\tilde f$ is the free-energy of the `ancestor' state, itself distributed
exponentially with a parameter $x(q-dq)$, etc. Assuming again that the barrier
between the states is related to the free-energy depth as pictured in Fig.
\ref{figjptree1}, one is led to surmise that
the trapping time $\tau_{\alpha,q}$ is still distributed as a power-law, but
with a $q$ dependent exponent:
\be
\rho_q(\tau) \propto_{\tau \gg t_o} \frac{t_o^{x(q)}}{\tau^{1+x(q)}}
\; .
\label{xq}
\ee
Note that $x(q)$ is an increasing function of $q$, which means that the smaller
the overlap between states,
the broader the distribution of time scales. In other words, `fast' processes
are deep down the tree.
The correlation function is now determined from {\footnote{With the convention
$\Pi_{M+1}\equiv 0$, and $q_{-1}=0$.}}
\be
C(t_w+\tau,t_w)= \sum_{j=0}^M q_j [\Pi_j(t_w+\tau,t_w)-\Pi_{j+1}(t_w+\tau,t_w)]
=
\sum_{j=0}^M [q_j-q_{j-1}] \Pi_j(t_w+\tau,t_w)
\; ,
\label{Cfulltree}
\ee
where $j$ labels the level of the tree from top to bottom, and
$\Pi_j(t_w+\tau,t_w)$ is the probability that no jump beyond the $j^{th}$ level
of
the tree has occured between $t_w$ and $t_w+\tau$. All the levels $j > M^*$
such
that $x(q_j)$ is larger than $1$ are equilibrated on microscopic ($\sim
t_o$) times. This means that the
corresponding $\Pi_j$ are zero as soon as $\tau \gg t_o$. This part of the tree
thus contributes to the
{\it stationary} dynamics, while the levels $j \leq M^*$ contribute to the
{\it aging} part
\be
C(t_w+\tau,t_w)= \sum_{j=0}^{M^*} q_j
[\Pi_j(t_w+\tau,t_w)-\Pi_{j+1}(t_w+\tau,t_w)] +
C_{\sc st}(\tau)
\; .
\ee
It is easy to show \cite{BD} that the short time $(\tau \ll t_w)$ decay of the
aging part of $C(t_w+\tau,t_w)$ behaves as in Eq. (\ref{Cshort}), with $x \to
x(q_{M^*})$, while the long time $(\tau \gg t_w)$ decay is decribed by Eq.
(\ref{Clong}) with $x \to x(q_0)$. Not surprisingly, the short time decay is
mostly sensitive to the fastest
part of the tree $j=M^*$, while the long time decay is governed by the slowest
processes $j=0$. Note that the experimental data is indeed such that the long
time $x(q_0)$ is smaller than the short time exponent $x(q_{M^*})$.

The presence of levels such that $x \simeq 1$ is very interesting from the
point of view of `$1/f$'-noise, for the following reason. If $x > 1$, the
corresponding contribution to the a.c. susceptibility is stationary
and behaves as $\chi''_{\sc st}(\omega) \propto (\omega t_o)^{x-1}$.
The noise spectrum is thus given by
\be
S(\omega) = \frac{2T}{\pi} \frac{\chi''(\omega)}{\omega} \propto \omega^{x-2}
\; .
\ee
So the leading contribution to the low frequency noise coming from these
equilibrated processes with $x>1$ comes from $x\simeq 1$ and scales as
$1/\omega$.
On the other hand, if $x < 1$, the contribution to the a.c. susceptibility is
aging
and behaves as $\chi''(\omega,t_w) \propto (\omega t_w)^{x-1}$. This
contribution is thus decaying with time,
but more and more slowly as $x$ approaches $1$ from below.
 Among these aging modes, those
which are the slowest to disappear correspond to  $x=1^-$, leading
again to a $1/\omega$ dependance. Within this picture,
 $1/f$ noise appears rather naturally; furthermore  one expects that this noise
should
generically exhibit some non-stationary contributions.

Another interesting aspect of the dynamics on a `multi-level' tree is its
response to temperature cycling.
It was suggested in Refs[\cite{saclayrev1,tempcycles2}] that the negative temperature
cyclings --
which reveal both `rejuvenation' in the intermediate, low temperature period,
followed by a perfect `memory' of the a.c. susceptibility --
point towards a hierarchical picture of phase space, where finer details are
progressively revealed as the temperature is lowered.
Using the fact that the whole curve $x(q)$ decreases when the
temperature is lowered, it is easy to account
for this phenomenon within the multi-level tree model \cite{BD}.

\vskip 0.5cm
$\bullet$ Energy or `entropy' barriers ?
\vskip 0.5cm
Before discussing the possible relation between this hierarchical picture and
real-space, droplet like descriptions, one should emphasize that in the trap
models
discussed above, aging is induced by the presence of energy barriers, the
crossing of which
becomes slower and slower as the temperature is decreased. Aging is
nevertheless also
present in models where there is no barrier crossing at all.
It was already noticed \cite{Eiop}
 that the {\sc sk} model at $T=0$ has a slow decrease in energy and never
reaches a stable configuration.
This same situation can be seen more clearly in the following model:
let us consider the case \cite{BarratM} where the
hopping rate $W_{\alpha \to \beta}$ is equal to zero if the state energy
$f_\beta$ is larger than $f_\alpha$, and equal to $W_0=(N t_o)^{-1}$
otherwise. This rule actually corresponds to the Glauber dynamics at zero
temperature. In the limit $N \to \infty$, the system never reaches the ground
state:
there always exists states of lower energy towards which the system can evolve
-- the number of these
`escape directions' however becomes smaller and smaller as time increases.
One can show that {\it independently} of the distribution of
energies $\rho(f)$ (but provided that these energies are independent), the
correlation function in this model is given by \cite{BarratM}
\be
C(t_w+\tau,t_w) = (q_{\sc ea}-q_0) \frac{t_w}{t_w+\tau} +q_0
\; .
\ee
Note that the decay of $C(t_w+\tau,t_w)$ is regular when $\tau \to 0$, in
contrast
with the above trap model and, as we shall see, with the generic mean-field
situation.
Slow dynamics in this model can thus be attributed to `entropic barriers', i.e.
the fact that paths
leading to smaller energies become more and more scarce as time increases. A
similar scenario holds in
the `Backgammon model' introduced by Ritort \cite{Ritort} and studied in
detail in Ref. [\cite{Backgammon}].
It is reasonable to expect that both energy and entropy barriers
should contribute to non-equilibrium dynamics in real systems.

\vskip 0.5cm

$\bullet$ Speculations about trees and clusters.
\vskip 0.5cm

What could be the interpretation of Eq. (\ref{Cfulltree}) in real space ?
Clearly, large $q$'s should be related to small clusters of reversed spins,
corresponding to `fast' processes, while small $q$'s
correspond to large clusters, or `slow' processes. Let us suppose that the
overlap between two
configurations can be written as a sum of contributions from `clusters' of
different linear scales $\ell$:
\be
q_{\alpha, \beta} = \frac{1}{L^d} \sum_{\ell=1}^L
\sum_{i_\ell=1}^{(\frac{L}{\ell})^d} q_\ell \Theta^\ell_{\alpha, \beta}
(i_\ell)
\; ,
\ee
where $L$ is the size of the sample, $q_\ell$ is the incremental contribution
to $q$ of the clusters of
size $\ell$; for fractal clusters \footnote{The
idea of fractal clusters of spins in spin-glasses dates back to
Ref [\cite{fractal}].} of dimension $d_f$ one expects $q_\ell
\propto \ell^{d_f}$. $\Theta^\ell _{\alpha, \beta} (i_\ell)$ is equal to zero
if
the states $\alpha$ and $\beta$ differ by the reversal of
the cluster of size $\ell$ in the `cell' labeled $i_\ell$ and equal to one
otherwise. Following the same speculative vein, one can write the two-time
correlation function as
\be
C(t_w+\tau,t_w) = \sum_{\ell=1}^L \frac{q_\ell}{\ell^d} \;
\Pi_\ell(t_w+\tau,t_w)
\label{Cell}
\ee
where $\Pi_\ell(t_w+\tau,t_w)$ is the average (over all the `cells' of size
$\ell$) fraction of
clusters which have not flipped between $t_w$ and $t_w+\tau$. The disorder
average is no
longer needed here since there is a {\it space} average over many independent
clusters.

Assuming that the
barrier heights are
distributed exponentially with a scale dependent parameter $x(\ell)$, we find
that the correlation function
will behave much as above in Eqs.(\ref{Cshort},\ref{Clong}). The parameter
$x(\ell)$ actually fixes the relation between energy scales $T/x(\ell)$
and length scales (see Refs.[\cite{MP,Balents}]). Taking $x(\ell) = T/(\Upsilon
\ell^{\theta})$ (as suggested by scaling arguments or replica calculations on
the problem
of pinned manifolds \cite{FH2,MP,Balents}), one finds that there exists a
characteristic length scale
$\ell^*$ such that $x(\ell^*)=1$, separating small length scales $\ell <
\ell^*$ -- for which equilibrium
is reached -- from large length scales where aging takes place. In other words,
\be
C(t_w+\tau,t_w) = \sum_{\ell=\ell^*}^L \frac{q_\ell}{\ell^d} \;
\Pi_\ell(t_w+\tau,t_w) +
C_{\sc st}(\tau)
\ee
with $C_{\sc st}(\tau)=\sum_{\ell=1}^{\ell^*} \frac{q_\ell}{\ell^d}
\Pi_\ell(\tau)$.
In particular, one has:
\be
q_{\sc ea} \equiv \lim_{\tau \to \infty} \lim_{t_w \to \infty} C(t_w+\tau,t_w)
=
\sum_{\ell=\ell^*}^L \frac{q_\ell}{\ell^d}
\; .
\ee
Assuming that \cite{FH} $\Upsilon \propto ({T_g-T})^\omega$, one thus finds
$q_{\sc ea} \propto (T_g-T)^\beta$, with $\beta = ((d-d_f)\omega)/\theta$.
Interestingly, as the temperature is decreased, there is an infinite sequence
of `glass transitions', where
all the length scales (in decreasing order) are progressively driven out of
equilibrium.
\vskip 0.5cm
$\bullet$ Conclusion
\vskip 0.5cm
Although there is still a lot of work to do to clarify the above picture
and make it consistent, the idea of modelling the dynamics of a complex system
through the motion of a point `particle' in a random potential is fruitful,
and actually used in many different contexts (structural glasses, protein
folding, etc.\cite{Phasespacemodels2}). The aim of the present Section was to
show that such models
can naturally lead to aging. Actually, the next Sections, devoted to an
analytical study of some mean-field models of spin glasses (which can also
be seen as models of diffusion in a random potential) will share many
similarities {\it and} important differences with the above discussion.

\section{Mean-field models of aging: analytical results}
\label{sectionmeanfield}

It took several years to realize that
mean-field models of spin-glasses, endowed with a suitable relaxational
dynamics (usually Langevin,
though Glauber is also possible), actually do capture some aging phenomena in
the glassy phase \cite{Bou92,Cuku1,Cukuri}.
They  thus provide a set of microscopical models where glassy dynamics and
aging effects can be studied analytically.
This Section will summarize the main results obtained in the recent years on
these mean-field models,
which have been most valuable in clarifying some of the basic theoretical
issues described in the
Introduction. The relation between these models and finite dimensional systems
is still very
much a matter of debate; we shall however postpone this discussion to Section
3.8 and the Conclusion.

The basic simplification occuring in mean-field models is that,
after averaging over the disorder and making the number of spins
large ($N \rightarrow \infty$), one obtains a set of {\it closed} equations for
the two-time correlation
and response functions, from which the energy and magnetization can also be
calculated. As we discuss below,
these equations
imply the existence of a critical temperature $T_c$, below which aging effects
appear, and the {\sc fdt} is
violated. On the other hand above $T_c$, the same equations allow for a {\sc
tti} solution, compatible with
{\sc fdt}.

Most mean-field dynamics studied so far have focused on models
which belong to the following family of spin glass Hamiltonians,
describing the interactions of $N$ continuous spins $\phi_i$, $i=1...N$:
\beq
E(\{\phi\})=  \sum_{r=1}^{\infty} F_r
\sum_{i_1<i_2, ...< i_{r+1}} J_{i_1,i_2, ..., i_{r+1}} \phi_{i_1} ...
\phi_{i_{r+1}}
\label{modelito}
\eeq
where the $J_{i_1,i_2, ..., i_{r+1}}$ are random Gaussian variables with
variance $N^{-r-1}$. Different
choices of $F_r$ lead to different models.

A quartic spin weight term can be added in order to make a `soft' version
of Ising spins, or  one can consider a spherical version, i.e.  $\sum_{i}
\phi_i^2=N$.
A popular choice is the  $p$-spin spherical model \cite{Crso}  defined by
$F_r=g \delta_{r+1,p}$,
plus a spherical constraint. In what follows  we shall mainly use
as an example
the spherical (or Gaussian) versions.
The same model (\ref{modelito}) can also be seen as the
potential energy of a point particle in a random potential, where the
$\phi_i$'s are the coordinate of the
particle's position in a $N$ dimensional space
\cite{toy}. In
this case, the energy is a Gaussian random potential, the correlations
of which are related to the $F_r$'s as
\bea
\overline{E(\{\phi \})E(\{ \phi' \})}=  N{\cal V} \left({1 \over N} \sum_i
\phi_i \phi'_i \right) \qquad
{\cal V} (x)=\sum_{r=1}^{\infty} \frac{F_r^2}{(r+1)!} x^{r+1}
\eea
where the overline means an average over the quenched disorder.

We shall in the following consider the dynamics to be modeled by a Langevin
equation:
\beq
\frac{ d\phi_i}{dt}= -\mu(t) \phi_i -\frac{\partial E}{\partial \phi_i} +
\eta_i(t) + h_i(t)
\eeq
where the white noises $\eta_i$ are mutually independent and of variance $2T$,
and $h_i(t)$ is a
time dependent external field. The `mass' term $\mu(t)$
is incorporated in order
to enforce the spherical constraint each time, or to model the presence of an
harmonic potential
in the case of a particle in a random potential, but may be set to zero in
other cases.

The correlations and response are defined as:
\beq
C(t,t')= \frac{1}{N}  \sum_{i=1}^N \overline{ \langle \phi_i(t) \phi_i(t') \rangle}
\; , \quad\quad\quad
 R(t,t')= \frac{1}{N}
 \sum_{i=1}^N
\left.\overline{\frac{\delta  \langle \phi_i(t) \rangle}{\delta
h_i(t')}}\right|_{h_i=0}
\label{defin}
\eeq
where the braces mean average over the thermal noises $\eta$.
There exist by now well established methods in order to obtain
the equations of motion for these systems in the large $N$ limit
(time being kept finite). The best known is to
introduce a dynamical field theory partition
function, average over disorder the partition function by
 using the fact that it is normalised \cite{cirano}, and compute
it for large $N$ by a saddle point method. This is the route which
was followed  originally
by Sompolinsky and Zippelius \cite{Sozi,So}. An alternative
derivation uses the cavity method. We refer the reader to the original papers
or to more recent textbooks \cite{Mepavi,Fihe}, and rather focus on the
solution of these equations. Starting the dynamics at time $t=0$
from a random configuration (chosen with a
uniform distribution in configuration space, corresponding to a quench from
infinite temperatures) the dynamical equations for the spherical or the
Gaussian case are found to be
\begin{eqnarray}
{\partial C(t,t') \over \partial t}
&=&
 - \mu(t) \, C(t,t') + 2 T \, R(t',t) 
\nonumber\\
& &
+\int_0^{t'} ds \; D(t,s) \; R(t',s)
+ \int_0^{t} ds \; \Sigma(t,s) \; C(s,t')\;,
\label{dyson1}
\\
{\partial R(t,t') \over \partial t}
&=&
 -\mu(t) \, R(t,t') +  \delta(t-t') +   \int_{t'}^{t} ds \;
 \Sigma(t,s) \; R(s,t')
\; ,
\label{dyson2}
\end{eqnarray}
where
\beqa
\Sigma(t,t')&\equiv& R(t,t') {\cal V}''[C(t,t')]
\; ,
\nonumber \\
D(t,t')&\equiv& {\cal V}'[C(t,t')]
\;.
\label{modecoup}
\eeqa
It is worth keeping in mind the limitations of the present approach. For
example, the simple form for $\Sigma,D$ in terms
of $C,R$ is a peculiarity of this class of models. More complicated forms in
which $\Sigma(t,t'),D(t,t')$
 are {\em functionals} of $C,R$ (with integrals involving $C$ and $R$ at
intermediate times)
rather than ordinary functions are obtained, for example, in the case of the
Sherrington-Kirkpatrick
model. So long as the functional dependency is only on $C,R$, one may however
expect that they can be treated with
the same methods. On the other hand, as soon as one introduces a Hamiltonian
with a finite number of neighbours per spin
(even for mean-field-like Hamiltonian such as Bethe lattice, or
random lattice systems), the dynamical
equations do not close on the two point correlation and response. One must then
introduce a whole hierarchy
of $k$ point correlations and responses,
which has not been investigated yet.

\subsection{Self-averageness and `universality'}

Equations (\ref{dyson1}), (\ref{dyson2}) and (\ref{modecoup}) are exact in the large $N$ limit
--- for times that do not
diverge with $N$. Furthermore, one can show using the same methods by which
they are
derived, that the correlations and responses are self-averaging with respect to
both
the thermal noise {\em and} the realization of disorder, again for
 times that do not diverge with $N$. Hence, we could well have omitted the
braces and the overbar in Eq. (\ref{defin}).
Non self-averageness in certain macroscopic quantities appears only for times
that
diverge with the system size, and in particular within the equilibrium
Gibbs-Boltzmann measure.

The  insensitivity of dynamics with respect
to the realisation of disorder is probably intimately related to the fact,
discussed in the next Section, that certain models
{\it without} quenched disorder show very similar {\em out of equilibrium}
behaviour, and hence can be studied by considering them as disordered.
The correspondence between the random  and
non-random version might however break
down for (divergent) times when non-self averaging features appear.

\subsection{The high temperature phase: two types of spin glasses}
\label{AvsB}

One expects that the above spin glass models converge fast towards
a paramagnetic equilibrium phase at high enough temperatures, where the system
has no long-term memory and obeys
{\sc tti} and {\sc fdt}. This can be seen
nicely from  a numerical study of the dynamical equations at relatively short
times, which will also be useful to identify  the qualitative behaviour at low
temperatures. In order to study the memory
of the system,
 we  plot the response functions $R(t,t')$ versus $t'-t$
 at different values of time $t$, $t=t_1,t_2,t_3,t_4$.
(These curves were obtained numerically from the  model model defined in 
Eq.(\ref{modelito}) with only
$F_2 \neq 0$. A simple numerical procedure
consists in discretizing time
evenly and iterating the dynamical equations which are causal. With some
extrapolation procedure on the mesh of the grid
one can reach safely times
of order $1000$ \cite{Frme1}.)


\begin{figure}
\centerline{\epsfxsize=9cm
\epsffile{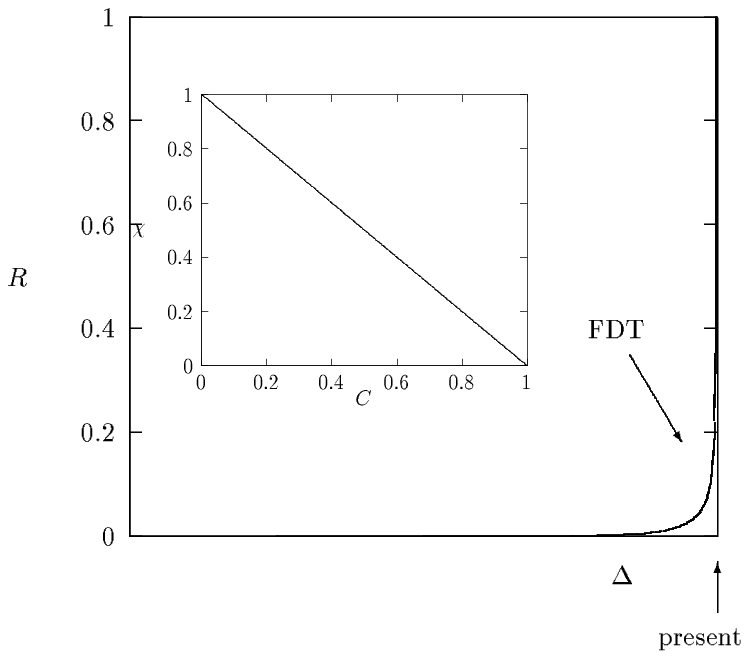}
}
\caption{The response $R(t,t')$ versus $t'-t$  at high temperatures
$T=1 >T_c$
for four total times $t=t_1,t_2,t_3,t_4$. The curves were
obtained from a numerical resolution of the dynamical equations (\ref{dyson1}),
(\ref{dyson2}) and
(\ref{modecoup})
in a $p=3$ spherical spin glass.
All four curves merge into one,
$R(t,t') = R(t-t')$. In the inset, $\tilde \chi(t,t')=\tilde \chi(t-t')$  vs.
$C(t,t')=C(t-t')$,
{\sc fdt} is satisfied and the slope  is just $-1/T$.
}
\label{highT1}
\end{figure}


Figure \ref{highT1} shows that the response at high temperatures
does not extend to the distant past and that it depends only on time
differences. The system achieves
equilibrium after a transient, and
eventually forgets the origin of times. Actually, the correlation function is
also {\sc tti}. Finally,
the $\tilde \chi$ vs. $C$ plot (see Section \ref{sectionfdt}) is a straight
line of slope $-1/T$ for all values of $C$
(see the inset in Fig. \ref{highT1}) showing that {\sc fdt} is satisfied.

Turning now to the analytical treatment of the equation, it is easy to show
that if {\sc tti} holds (i.e. if the two unknown functions of two variables
$C(t_w+\tau,t_w)$ and $R(t_w+\tau,t_w)$ actually depend only
on $\tau$), then the second equation (\ref{dyson2}) is a consequence of the
first (\ref{dyson1}) provided
the {\sc fdt} is satisfied. We are thus left
with a single function $C(\tau)$ in Eq. (\ref{dyson1}), and the kernels
$\Sigma$ and $D$ are related by
$d D /d \tau = -\Sigma/T $. Using the finite
extent of the memory we can safely send the initial time to  $-\infty$, to
obtain:
\beq
\frac{dC(\tau)}{dt}
=
-\mu_\infty C(\tau) +
\frac1T \int_{-\infty}^{\tau} d \tau' \, {\cal V}'[C(\tau-\tau')]
\frac{\partial C(\tau')}{\partial \tau'} \; , \qquad\qquad C(\tau=0)=1
\label{mcthT}
\eeq
where
\be
\mu_\infty = \mu(t \to \infty) - {\cal V}'[1]/T
\; .
\label{Gogo}
\ee
The above equation is valid as long as $C(\tau)$ decays to zero in the long
$\tau$ limit.

Equation (\ref{mcthT}) is basically the general `schematic' Mode-Coupling
equation for the density
correlations in a supercooled liquid above the dynamical transition temperature
introduced by Leutheusser \cite{leut}, G\"otze and  others \cite{Go} as a
model for the ideal glass transition. The only difference lies in the fact that
the Mode-Coupling equations
also possess an `inertial' term ${\partial^2_\tau C(\tau)}$. This coincidence
will be further discussed in
Section \ref{modecouplingglass}.

The behaviour of the solution to these equations
when one lowers the temperature depends upon the
structure of the disorder, i.e. upon the function ${\cal V}$. It turns out that
there are two broad classes of mean-field spin glasses,
characterized by rather different behaviours.
The point is that as one lowers the temperature, there
appears a critical point $T_c$ at which the decay behaviour of $C(\tau)$
shows a marked change.
For `discontinuous' models, $C(\tau)$  does no longer decay to zero,
but rather to a finite value $q_{\sc ea}>0$ (which is also called the non
ergodicity parameter $f$ in the context of glasses).
For `continuous' models, $C(\tau)$ still decays to $q_{\sc ea}=0$ at $T_c$ but
with critical slowing down. Below $T_c$, $q_{\sc ea}$ grows continuously from
zero \footnote{These two types
are also called, respectively, A and B in the context of the Mode Coupling
theory \cite{Go}. One should keep in mind
that we speak here of dynamical phase transitions, and
this terminology is not related to the Ehrenfest classification of equilibrium
phase transitions.}.

\vskip 0.5cm
$\bullet$ `Discontinuous' models
\vskip 0.5cm

The simplest prototype, which we discuss here, is the pure spherical $p$-spin
model with $F_r= g \delta_{r+1,p}$.
Another example is provided by the problem of a particle in a random potential
with short range correlations, i.e. when
${\cal V}(x)$ decays sufficiently fast for large $x$'s.


\begin{figure}
\centerline{\epsfxsize=9cm
\epsffile{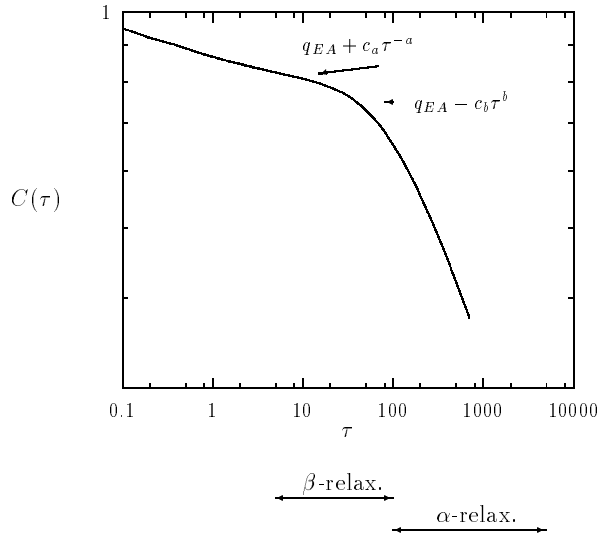}
}
\caption{The correlation for the same model as in Fig. \ref{highT1} at high
temperatures close to the transition
$T \stackrel{>}{\sim}T_c$}
\label{revfigChighT2}
\end{figure}


The solution, studied in Refs.[\cite{Go,Crhoso,Kiho1}] behaves as follows.
Above the critical temperature $T_c$, the correlation $C(\tau)$ decays to zero
at large $\tau$.
Slightly above $T_c$, the correlation already starts developing a plateau
at $C \sim q_{\sc ea}$ before eventually decaying to zero, as shown in Fig.
\ref{revfigChighT2}.
The length of the plateau increases as a power law of $T-T_c$ when
 the temperature gets closer to $T_c$.
The details of how $C(\tau)$ first decays towards the
plateau and then departs from it has been one of the most studied aspect of the
{\sc mct} in the context of glass
forming liquids, since these features can be
directly tested experimentally using a variety of techniques. One finds that:
\begin{eqnarray}
\begin{array}{rcll}
C(\tau) &\sim & q_{\sc ea} + c_a \tau^{-a} \qquad & C {> \atop \sim} q_{\sc ea}
\; ,
\\
C(\tau) &\sim & q_{\sc ea} - c_b \tau^b  \qquad & C {< \atop \sim} q_{\sc ea}
\; ,
\end{array}
\label{HighTmct}
\end{eqnarray}
where the exponents $a,b$ are related by
\beq
\frac{\Gamma^2[1+b]}{\Gamma[1+2 b]}
=
\frac{\Gamma^2[1-a]}{\Gamma[1-2 a]}
=
\frac{T_c}{2} \,
\frac{ {\cal V}'''(q_{\sc ea})}{({\cal V}''(q_{\sc ea}))^{3/2}}
\; .
\label{mctabHT}
\eeq
Clearly, the equilibration time for a  system  that is quenched to $T > T_c$ is
at least as large as the time of the plateau $\tau(T)$. A formulation in terms
of time-differences (\ref{Gogo}) is thus valid only when the waiting time is
much larger than the $\alpha$-relaxation time $\tau(T)$.
For times that are smaller than $\tau(T)$, one has to go back
to the two-time equations --- this will be the case throughout the
low-temperature
phase.

\vskip 0.5cm
$\bullet$ `Continuous' models
\vskip 0.5cm

The class of `continuous' spin glasses contains the more usual
case of the {\sc sk} model \cite{Sozi}
(which is however not described by an equation of
the type  (\ref{mcthT})). Among the systems we are discussing is
the case of a particle in a long-range correlated random potential \cite{Kiho2}
(${\cal V}(x)$ decaying as a power law), or some spin systems such as
 a mixture
between $p=2$  and $p=4$ interactions\cite{Theo2}: $F_r=g \delta_{r+1,2}+g'
\delta_{r+1,4}$.

Again there is a critical temperature $T_c$ above which
the correlation $C(\tau)$ decays to zero at large $\tau$. The main difference
with the previous case is the absence of the plateau
structure around $q_{\sc ea}$ for $T$ slightly above $T_c$, which is obviously
related to the fact that in this case the Edwards-Anderson order
parameter departs continuously from zero when
one decreases the temperature through $T_c$.

\subsection{Low temperatures: Weak long term memory and weak ergodicity
breaking}

Let us now discuss what happens below $T_c$. Again, we first show
the numerical solution of the full causal dynamical equations
(\ref{dyson1}), (\ref{dyson2}) and (\ref{modecoup}), and look at the
same plot as before, namely the response functions $R(t,t')$ versus $t'-t$, for
different values of  time $t$, $t=t_1,t_2,t_3,t_4$. Figure \ref{revfigRlowT}
shows that:


\begin{figure}
\centerline{\epsfxsize=9cm
\epsffile{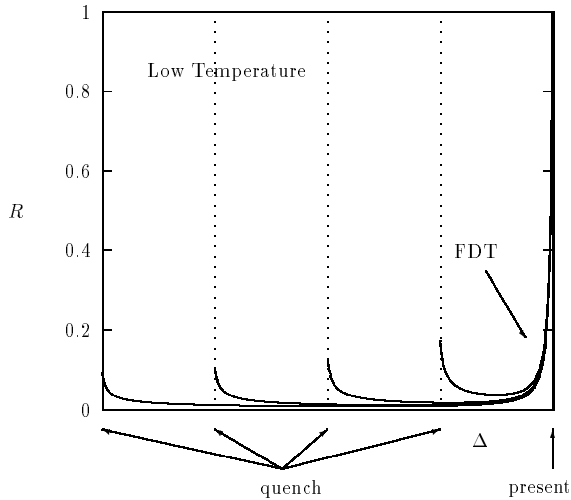}
}
\caption{The response $R(t_i,t')$ in terms of $t'-t_i$
at a low temperature $T<T_c$ for the same model as in Fig.\ref{highT1}.
For each curve $t_i$ is $t_1=100$, $t_2=200$, $t_3=300$, $t_4=400$,
respectively. `Quench' corresponds to $t'=0$ while `present' to 
$t'=t_i$.
}
\label{revfigRlowT}
\end{figure}



\begin{figure}
\centerline{\epsfxsize=10cm
\epsffile{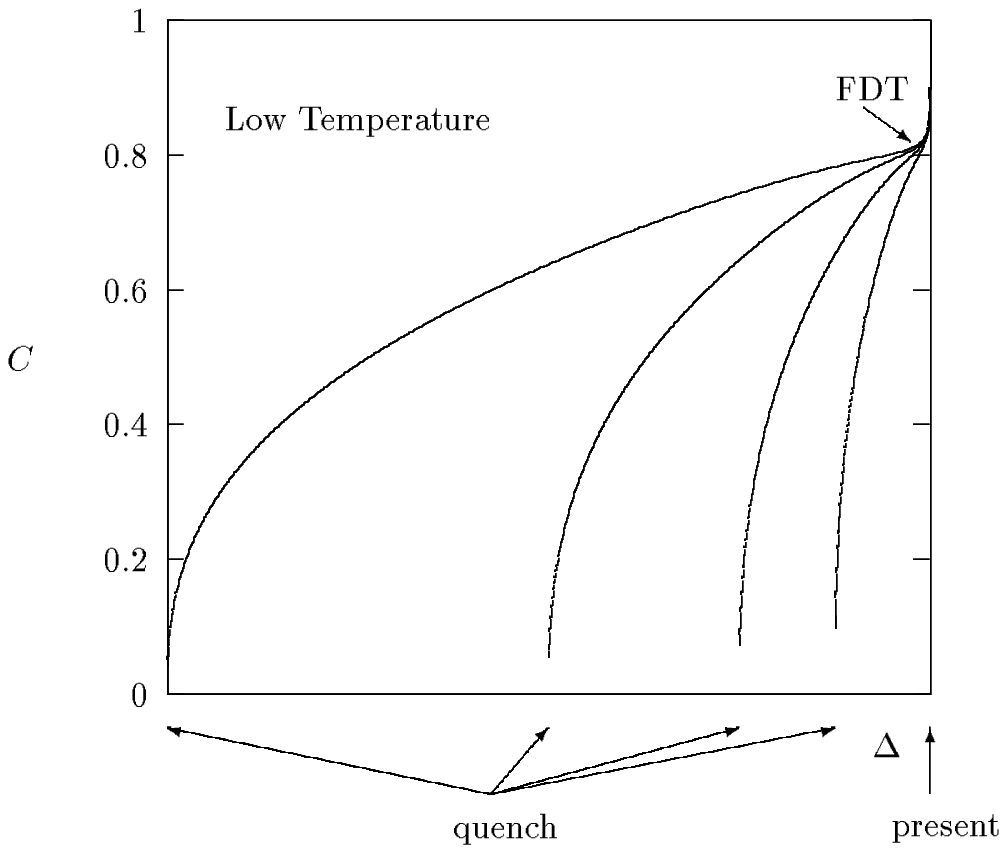}
}
\caption{The correlation function $C(t_i,t')$ vs $t'-t_i$ at
low temperatures for the same model as in Figs \ref{highT1}
and \ref{revfigRlowT}. From left to right $t_i=400,200,100,50$.
}
\label{revfigClowT}
\end{figure}


--  The system has a strong response to pertubations in the immediate past that
is quite similar to the high temperature response.

-- However, a long tail extending down to the quench time $t'=0$  has now
appeared.

The total area under the response curves,
$\int_0^t ds \, R(t,s)$  approaches at large times a  finite
limit, which is equal to the linear susceptibility $\tilde \chi$
to a constant field. Part of this area is already
given by the peak to the right of Fig. \ref{revfigRlowT}, which is the
high-frequency, stationary
contribution to the susceptibility.
It turns out however that the area below the long-time tail also gives a
non-zero contribution -- the memory to the
distant past is substantial and can never be neglected:
\beq
\forall t^*
\;\;\;\;\;\;\;\;\;\;
\lim_{t\to\infty}
 \int_{t-t^*}^t ds R(t,s)
<
\lim_{t\to\infty}
 \int_{0}^t ds R(t,s)
\; .
\eeq
This can be hinted from the simulations. As we shall see, it can also be
derived
from the dynamical equations.

At this stage it would  seem that  in order to solve for the correlations and
responses at large times we  need to know the complete solution at {\em all}
times, because the memory kernels
(which involve the response) have nontrivial contributions from all the past.
If this were the case, the problem of finding an asymptotic solution
for long times would be hopeless from the analytical point of
view!

Fortunately, something remarkable happens within the models considered here:
even though the area under the
long-time tails of the response remains finite, the height of the tails
themselves
 tend to zero as we consider larger times. More precisely, the
integrated response at time
$t$ to a signal between the initial time $s=0$ and any
{\em finite} time $s=t^*$ tends to zero at large $t$:
\beq
\lim_{t \rightarrow \infty} \int_0^{t^*} ds \, R(t,s) =0
\label{wltm}
\eeq
for any fixed $t^*$. The same can be said about the integral of $R \times G$
where $G$ is
any finite function, such as the memory kernel $\Sigma$.
This behaviour was described as `weak long term memory' in Ref. [\cite{Cuku1}]:
the memory tends to be
`weak' for
any finite interval of the past, but is strong when integrated over the whole
past
\footnote{
Not every system will
satisfy this condition; on the contrary, systems
that remember their initial transients are much harder
to treat and how to deal with their dynamics is still
a fully open problem.
}. This is
close in spirit to the `weak ergodicity breaking'
property defined in Section (\ref{WEB}):
 a perturbation lasting for any finite duration will eventually be forgotten.
Therefore the long time dynamics will decouple from the initial
(non universal, and out of control) transients.

The evolution of the
two-time correlation function also has (at least) two distinct regimes, as
shown in Figs. \ref{revfigClowT}  and \ref{revfigClowT2}.
For times such that $\tau=t-t'$ is small all curves merge and one has {\sc
tti}.
(These are times close to `present' in Fig. \ref{revfigClowT}
and to the left in Fig. \ref{revfigClowT2}.)
The corresponding decay is `fast' (see footnote$^b$ above) in the regime where
$C$
drops from $1$ at equal times to the plateau value $q_{\sc ea}$, defined in
Eq.(\ref{qEAdef}).  However, when $C$
decays below $q_{\sc ea}$ it does it in a manner
which depends on both $t$ and $\tau$. This subsequent decay is `slow'.

This suggests  (and the analytical calculation later
confirms) that one can perform, for both the correlation and response
functions,
a decomposition into a stationary and an aging part, similar
to the one introduced above for the description of aging experiments
(see Figs. \ref{revfigchiexp} and \ref{revfigmexp}):
\begin{eqnarray}
R(t_w+\tau,t_w) &=& R_{\sc ag}(t_w+\tau,t_w) + R_{\sc st}(\tau)
\; ,
\nonumber \\
C(t_w+\tau,t_w) &=& C_{\sc ag}(t_w+\tau,t_w) + C_{\sc st}(\tau)
\; .
\label{aging_decomposition2}
\end{eqnarray}

Defining as in Eq.(\ref{qEAdef}) the Edwards-Anderson order parameter: $q_{\sc
ea} =$
 $\lim_{\tau \to \infty}
\lim_{t_w \to \infty}$ $ C(t_w+\tau,t_w)$, the stationary (and thus the aging)
parts are defined by:
\be
C_{\sc st}(\tau) \equiv \lim_{t_w\to\infty} C(t_w+\tau,t_w) -q_{\sc ea}
\qquad
R_{\sc st}(\tau) \equiv \lim_{t_w\to\infty} R(t_w+\tau,t_w) \label{eqcontr}
\ee
 It turns out that the stationary parts, which are by definition {\sc tti},
also satisfy the {\sc fdt},
$R_{\sc st}(\tau) = -\frac1T d C_{\sc st}(\tau)/d\tau$. Finally, note that the
above definitions imply:
\bea
\begin{array}{rclrcl}
\lim_{t_w\to\infty} C_{\sc ag}(t_w+\tau,t_w) &=& q_{\sc ea}
&
\lim_{t_w\to\infty} R_{\sc ag}(t_w+\tau,t_w) &=& 0
\\
\lim_{\tau \to \infty} C_{\sc st}(\tau) &=& 0
&
C_{\sc st}(0)&=& 1-q_{\sc ea}
\\
 \lim_{\tau \to \infty}R_{\sc st}(\tau)&=& 0
&
R_{\sc st}(0) &=& 1
\end{array}
\end{eqnarray}

\vspace{.5cm}
\subsection{Low temperature solution of the dynamical equations}
\label{solution}

\vskip 0.5cm
$\bullet$ General strategy.
\vskip 0.5cm
An asymptotic solution to the dynamical equations was first found
in Ref. [\cite{Cuku1}] in the $p$-spin spherical model and then generalised
to various situations \cite{Frme}$^-$\cite{Cule}$^,$\cite{Cukule,BBM}.
 It has been described in detail in several
papers \cite{Cuku2,Cule,BBM}. We restrict here to the essential
 assumptions and the main ideas allowing one to find the low temperature
solution of the
dynamical equations (\ref{dyson1})  and (\ref{dyson2}).

The solution strongly relies on the
weak long term memory assumption. It
is asymptotic in the sense
that it holds only at large times $t_w$: the transient
effects are much more complicated, and have not been
studied yet.
The method of solution we shall outline here only determines
the time-dependences  in the aging regime up to a reparametrization
in time $t \rightarrow h(t)$.
Thus, one obtains a
 family
of solutions for the aging part, related to one another by time
 reparametrisations:
\be
C_{\sc ag}(t,t') \rightarrow C_{\sc ag}(h(t),h(t'))
\qquad
R_{\sc ag}(t,t') \rightarrow \left( \frac{dh(t')}{dt'} \right) \, R_{\sc
ag}(h(t),h(t'))
\; .
\ee
(Note that the presence of the factor $dh(t')/dt'$ comes from the fact that
it is the integral of $R$ over time, rather than $R$ itself, which is the
physical
quantity).
The `selection problem' of determining the actual function $h$ chosen by the
system is still
an open one that requires
 the  matching of the  regimes of short and long time differences.

The starting step consists in proposing asymptotic forms for the aging part of
the correlation
function in the two-time plane valid for large $t_w+\tau$ and $\tau$.
As we mentioned in  (\ref{Timesectors}), a possible form for the correlation
is:
\beq
C_{\sc ag}(t_w+\tau,t_w) \sim \sum_i {\cal{C}}_i \left[
\frac{h_i(t_w)}{h_i(t_w+\tau)} \right]
\label{domingo}
\eeq
where each term will vary in separate time sectors, defined by taking the times
to infinity with
 $0 < {h_i(t_w)}/{h_i(t_w+\tau)} < 1$
(see Eq.(\ref{domi})).
The asymptotic form of the correlation is given then by the knowledge of
$ {\cal{C}}_i $ and $h_i$.
This form is meant to
represent the correlation only in the limit of large times, and it need
not be unique.
One may wonder whether it exhausts all possibilities.

This question can be answered with the following construction\cite{Cuku2}.
Consider the configurations at three
large times   $t_{min} \leq t_{int} \leq t_{max}$, and the corresponding
correlations
$C(t_{max},t_{min})$, $C(t_{max},t_{int})$ and $C(t_{int},t_{min})$.
Using the fact that the correlations decrease with time-separations,
one can show that in the limit of large times
the  three correlations must be related by
\beq
C(t_{max},t_{min})= f[ C(t_{max},t_{int}), C(t_{int},t_{min})]
\; 
\label{tritri}
\eeq
where the function $f$ defines the geometry of the triangles described by the
trajectory of
the system in phase space. Now, it is easy to see that $f$ is an associative
function,
and one can classify all the possible forms of an associative, monotonical
function \cite{Cuku2}  using
elementary group theory.

This in turn leads to a classification of all the possible two-time scalings as
follows.
One considers  the special (fixed point) values of the correlation $q_0, \; 
q_1, \dots, \; q_k \equiv q_{\sc ea}, \; q_{k+1} \equiv C(t,t)$ defined by
$f(q_i,q_i)=q_i$. If the number of fixed points is finite, one  can prove that
the correlation can be written
for large times as (\ref{domingo}) with $q_i= \sum_{j<i} {\cal {C}}_j(1)$.
The intermediate values of correlation between two fixed points constitute a
`correlation scale', within which
only one term of (\ref{domingo})
is non-constant.
 Triangles whose  sides  belong to different scales (e.g.
$C(t_{max},t_{int})<q_i$ ,  $C(t_{int},t_{min})>q_i$) are isosceles with
$C(t_{max},t_{min})= \min[ C(t_{max},t_{int}), C(t_{int},t_{min})] $: there is
ultrametricity
\cite{ramtouvir} between correlation scales.
It may also happen that the fixed points $q_i$ form a continuum within a
certain range $(q_a,q_b)$. In that
case we have, for any two correlations
within this range:
\beq
C(t_{max},t_{min})= \min[ C(t_{max},t_{int}), C(t_{int},t_{min})]
\;\;\;\;\;\;\;\;\;\; {\mbox{as}} \  \ t_{min} \to \infty
\; .
\label{ultra}
\eeq
Correspondingly, the correlation  has to be represented in that case as   a
limit of   a continuous
sum of infinitely many scaling functions ${\cal C}_i$ with vanishing weight.
Note that the above construction  does not rely on any concrete model and is
not restricted  to mean-field.

It is interesting to see that this form of ultrametricity in the correlations
 appears in a very natural
way within out of equilibrium dynamics.
It is  already present in the simplest form in the case of domain
growth: if we consider three very large times such that
$C(t_{max},t_{int})>m^2$
and $C(t_{int},t_{min})< m^2$  (where $m$ is the magnetization),
then $C(t_{max},t_{min})= \min[ C(t_{max},t_{int}), C(t_{int},t_{min})] $.
Ultrametricity between correlations larger and smaller
than $m^2$ expresses the fact that the time scale
corresponding to relaxation due to domain wall motion becomes infinitely
larger than the one corresponding to thermal fluctuations within each domain.
It is however not clear that there are finite dimensional models for which a
`full' ultrametric form
(\ref{ultra}) holds for {\em any} two values of the correlations.

Having discussed the possible asymptotic forms of the correlations,
the next step is to make an Ansatz for the response function. One can write
without loss of generality:
\beq
R(t,t')= \frac{X(t,t')}{T} \frac{\partial C(t,t')}{\partial t'}
\; 
\label{quaqua}
\eeq
($t\geq t'$). The Ansatz now consists in proposing that for large $t,t'$, $X$ only depends on
time through $C$:
\beq
X(t,t')=X[C(t,t')]
\eeq
or, in other words, that the parametric plots $\tilde \chi$ vs $C$ in Section
\ref{sectionfdt}
converge to a limit curve as $t_w \to \infty$.
The  asymptotic form for the response is thus obtained from that of the
correlation
through the introduction of a certain function $X[C]$. At this point one
substitutes the
 above Ans\"atze for the correlation and response
 functions into the dynamical equations. In this way one determines

 {\em i)} $X[C]$ (which contains the information on aging of the response)
 and

 {\em ii)}
 $f$, or, equivalently, the number of terms in (\ref{domingo})  and the
respective
 ${\cal{C}}_i$.

 {\em  iii)} One should in principle also obtain the $h_i$, but this requires
to
solve the selection problem discussed above.

The corresponding computations can be rather lengthy. We shall not detail them
here, but rather
present the results obtained when one applies this technique
 to the mean-field models considered above.

\vskip 0.5cm
$\bullet$ The case of discontinuous models.
\vskip 0.5cm

For low temperatures   $T<T_c$
the correlation is given by Eq.(\ref{domingo}) with only {\em one} aging sector
(i.e. one function $h(t)$) plus the {\sc tti} scale.
 It reads
\be
C(t_w+\tau,t_w) = C_{\sc st}(\tau) + C_{\sc ag}
\left(\frac{h(t_w)}{h(t_w+\tau)} \right)
\; .
\label{C1}
\ee

The memory properties of the system are controlled by
$X[C]$, which takes a particularly simple form:
\beqa
X[C] = 1 \;\;\;\;\;\;\;\;\;\;\;\;  &{\mbox{for}}&  C>q_{\sc ea}     \;\;\;\;
\;\;\;\;  {\mbox{({\sc fdt})}} \nonumber \\
X[C]  =X<1  \;\;\;\;  &{\mbox{for}}&  C<q_{\sc ea}
\; .
\eeqa
$X$ is a  positive, temperature-dependent  number, constant
throughout the aging regime.
The fact that $X$ is constant is not assumed {\em a priori}, but comes out from
the equations of motion. A possible explanation for this fact
is found when interpreting $T^{\sc eff} = T/X$ as
an `aging temperature': it  means that degrees of freedom of comparable
 frequencies are mutually thermalised \cite{Cukupe}. Note that
at the glass transition, the aging temperature is equal to the bath
temperature. The ratio $T^{\sc eff}/T$ {\em increases} with decreasing bath
temperature; in particular, $T^{\sc eff}$ remains non-zero at zero bath
temperature.

The behaviour of the correlation around the plateau $q_{\sc ea}$ provides  the
low-temperature
extension of the one already encountered at high temperatures.
Taking the limit of large $t_w$ {\it before} taking the limit of large $\tau$
one only explores the
 approach of the correlation to the plateau $q_{\sc ea}$. This decay is still
given by a power law
 characterised by a temperature-dependent exponent $a$, as was the case for
$T>T_c$. However, the subsequent
 departure from $q_{\sc ea}$ is $t_w$-dependent and  characterized \cite{Cule}
by another temperature-dependent exponent $b$:
\begin{eqnarray}
\begin{array}{rcll}
C(t_w+\tau,t_w) &\sim & q_{\sc ea} + c_a \tau^{-a}
\qquad
&
C {> \atop \sim} q_{\sc ea}
\\
C(t_w+\tau,t_w) &\sim & q_{\sc ea} - c_b \left(\frac{\tau}{{\cal T}_w}\right)^b
\qquad
&
C {< \atop \sim} q_{\sc ea}
\end{array}
\label{hh}
\end{eqnarray}
where ${\cal T}_w$ is an effective waiting time, defined as
${\cal T}_w=h(t_w)/h'(t_w)$. (In
the case in which  $h(t)$ is a simple power law, one has ${\cal T}_w \propto t_w$.)
The exponents $a,b$ have precisely the same
meaning as the exponents $a,b$ defined in Section (\ref{experiments}), and are
in this case related by
\beq
X \frac{\Gamma^2[1+b]}{\Gamma[1+2 b]}
=
\frac{\Gamma^2[1-a]}{\Gamma[1-2 a]}
=
\frac{T}{2} \,
\frac{ {\cal V}'''(q_{\sc ea})}{({\cal V}''(q_{\sc ea}))^{3/2}}
\label{mctabLT}
\eeq
with $q_{\sc ea}$ given by Eq.(\ref{qEAdef}).

\vskip 0.5cm
$\bullet$ The case of continuous models.
\vskip 0.5cm

As regards  the temporal behaviour of the correlations in the stationary
regime,
the approach to the plateau at $q_{\sc ea}$ is also given by
a power law with a temperature-dependent exponent $a$.
The situation within the aging regime is however more complicated
than in the case of discontinuous models.
The aging part of the correlations satisfy the full ultrametric triangle
relations (\ref{ultra}):
\beq
f(C_1,C_2)=\min[C_1,C_2]
\eeq
if at least one of $C_1,C_2$ is smaller than $q_{\sc ea}$.
A representation like (\ref{hh}) for the escape from the plateau is not
possible for these models unless one makes the
exponent $b$ waiting-time dependent.

For purely continuous models the function $X[C]$ is not a constant and,
remarkably, coincides with the function $x(q)$
in the replica treatment of equilibrium\footnote{The dynamical $X[C]$ is much
easier to obtain numerically than its static counterpart
because one does not have to equilibrate.
The  numerical confirmation of the analytical form of $X[C]$ for the {\sc sk}
model is astonishingly good \cite{Bacukupa,ferraro}. }.
 This and other coincidences between out of equilibrium dynamics and statics of
continuous models have  escaped any
kind of physical understanding so far. (For some recent work in this direction,
see Ref. [\cite{sknum}].)

\subsection{Generalization to several coupled modes --- the case of spatial
dependence}

New physical insights appear when we consider the
 generalization to several coupled models, and, in particular,
to mean-field cases in which there is also spatial dependence.
In such cases we have to deal with several correlations $C_{k,k'}(t,t')$ and
responses
$R_{k,k'}(t,t')$, where the indices $k,k'$ refer to different modes, for
example they can
represent spatial positions $x_k$ or Fourier components.

The construction of an Ansatz proceeds as before\cite{Cukule}. One has to add now a
prescription for
the {\em long-times} relationship between different correlations. Choosing
one particular correlation function $C_{0,0}$ as an effective `clock', one may
look for solutions of the form:
\beq
C_{k,k'}(t,t')={\cal F}_{k,k'}[C_{0,0}(t,t')]
\label{ans}
\eeq
 with ${\cal F}_{k,k'}$ to be determined.
One also introduces the fluctuation-dissipation violation factors defined by
\beq
R_{k,k'}(t,t')= \frac{X_{k,k'} [C_{k,k'}(t,t')]}{T} \frac{\partial
C_{k,k'}}{\partial t'}
\label{ans1}
\; .
\eeq
Interestingly, it turns out that the Ansatz closes with two extreme
possibilities:

{\em i)}  $X_{k,k'}\neq 0$ for $k \neq k'$,  and $X_{k,k}=X_{k',k'}$ at equal
times;

{\em ii)} $X_{k,k'} \rightarrow 0$ for $k \neq k'$, and
$X_{k,k},X_{k',k'}$ possibly different.

This can be understood as a property of partial thermalisation \cite{Cukupe}:
remembering
that $T_{k}\equiv T/X_{k,k}$ is an effective temperature, in the case {\em i)}
the subsystems have
$T_{k}=T_{k'}$ at corresponding time scales, while in case  {\em ii)} the
subsystems
have zero cross response and $T_{k}$ may be different from $T_{k'}$.

A  mean-field case with many modes is obtained  when one studies the
 dynamics of a random manifold in a disordered medium within the Hartree
approximation
\cite{Cukule}.
One has all the time scalings described so far for each mode $k$, plus
dynamical scalings
in terms of $k$ and times. It turns out that a solution satisfying (\ref{ans})
and (\ref{ans1}) appears naturally  in that case.

\begin{figure}
\centerline{\epsfxsize=10cm
\epsffile{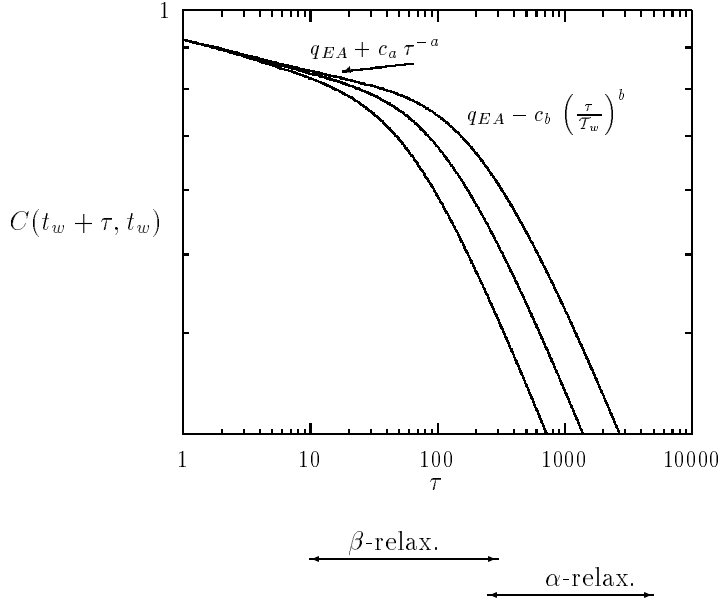}
}
\caption{The correlation $C(t_w+\tau,t_w)$ as a function of $\tau$ at a low
temperature $T < T_c$.
$t_w$ is $t_w=50,100,200$, respectively.}
\label{revfigClowT2}
\end{figure}


\vskip 0.5cm
\subsection{Speculations on the `effective' age function $h(t)$}
\label{onh(t)}
\vskip 0.5cm

Although the above solution describes some general features of the low
temperature,
aging regime of mean-field spin glass models, it is still,
even in the simpler discontinuous case, incomplete. As mentioned above,
the dynamical equations become, in the asymptotic ($t \to \infty$) limit,
invariant under
any monotonous time reparametrisation $t \to h(t)$. The function $h(t)$ can in
principle
only be determined through matching  with the early ($\tau \sim t_o$) solution,
or by a
numerical
solution of the two-time equations. For the spherical  $p$-spin problem, the
latter procedure
suggests
that $h(t)$ is close
to a power-law \cite{Cuku1}, in other words that the aging part of the
correlation function
$C(t_w+\tau,t_w)$
is a function of the ratio $\tau/t_w$. In principle, $h(t)$ could be
any other function of time, for example:
\beq
h(t)=\exp\left[\frac{1}{1-\mu} \left(\frac{t}{t_o} \right)^{1-\mu} \right]
\quad \mbox{or}
\quad
h(t)=\exp \left[\log^\nu(\frac{t}{t_o}) \right]
\eeq
in which case, the effective `age' ${\cal T}_w$ appearing in, e.g. (\ref{hh})
is
\beq
{\cal T}_w \equiv \frac{h(t_w)}{h'(t_w)} = t_w^\mu t_o^{1-\mu} \quad \mbox{or}
\quad
\frac{t_w}{\nu \log^{\nu-1}(\frac{t_w}{t_o})}
\; .
\eeq
Except in the cases $\mu=1$ or $\nu=1$, where $h(t)=t/t_o$,
these more complicated forms lead to an explicit dependence of the effective
age ${\cal T}_w$ on the microscopic time scale $t_o$. (More precisely, when
$\mu < 1$ (or $\nu > 1$),
the effective age ${\cal T}_w$ is much smaller than $t_w$, a situation called
`subaging' in Ref.[\cite{saclayrev2}].)

The simplest scaling situation would be that the value of $t_o$ becomes
irrelevant on
the experimental time scale where $t_w, \tau \gg t_o$. This would lead
immediately
to `full' aging, ${\cal T}_w \propto t_w$. Naive scaling
can however break down in some cases \cite{Barenblatt}, which means that the
value of $t_o$ is important even in the $t_w \to \infty$ limit -- in other
words that the effective age ${\cal T}_w$ depends on $t_o$. Such is the case of
systems with logarithmic domain growth {\em e.g.}
the random field Ising models, where ${\cal T}_w \propto t_w
\log(\frac{t_w}{t_o})$ (`superaging'). Signs that
subaging may also
happen in the mean-field models considered here can be found in
\cite{Hornermatching}.
We feel that the determination of $h$  is
one
major unsolved issue in mean-field dynamics. Correspondingly, the
same ambiguity remains from an experimental point of view:
a detailed analysis of the {\sc trm} reveals small but systematic
deviations from `full' aging. A scaling function $h(t)$ of the above form, with
$\mu=0.97$
or $\nu \sim 2$, does a better job at fitting the experiments
\cite{saclayrev2} (see Figs. 3.b and 3.c. in Ref.[\cite{saclayrev2}]).
 However, these deviations might  alternatively be interpreted as  being
`interrupted aging'  ({\em i.e.} equilibration in
a long but finite time)
due to finite field \cite{St,ustocome} or finite size \cite{Bou92,JPEVJH}
effects.

At any rate, note that a pure $\tau/t_w$ behaviour would rule out the existence
of many time sectors with
full ultrametricity, and would leave a scaling as in the discontinuous models
as the only possibility.

\subsection{Out of equilibrium versus `equilibrium on diverging time scales'}
\label{statics}

Dynamical studies of mean-field spin glass models
in their low temperature phase are not new. They started
more than 15 years ago with the work of Sompolinsky and Zippelius
\cite{Sozi,So}
and were subsequently applied to many other problems. The above description,
which focuses onto out of equilibrium dynamics, has
followed a very different route -- which was only found recently \cite{Cuku1}.
However, as all these works address similar issues, a comment
on their relationship is in order. For lack of space we shall not
be able to present in any detail the former approaches, but these
are by now well documented \cite{Sozi,So,Biyo,Mepavi,Fihe}. We rather want to
stress the
conceptual differences with the present approach.

Consider a mean-field spin glass model below $T_c$. The two times
dynamical equations (\ref{dyson1}) and (\ref{dyson2}) are exact equations
relating
the correlation and response in the mean-field models.
They describe the behaviour of the system  at  times that do not diverge with
$N$. Even when speaking of `long
times' within this framework,  one really means:
\beq
 \lim_{t,t' \rightarrow \infty} \lim_{N \rightarrow \infty}
\; .
\label{tN}
\eeq
Obviously for finite $N$ the spin glass will equilibrate in a finite
time. There exists an equilibration time $t_{\sc erg} (N)$, such that for $t_w$
much larger than $t_{\sc erg}(N)$ the system is equilibrated, which means that
the configurations are sampled with a frequency proportional to their
Boltzmann-Gibbs weight. The correlation and response then
become time translational invariant, and related by the {\sc fdt}.
At large $N$ the  equilibration time diverges.

The approach of Ref. [\cite{So}] starts from the very same dynamical equations
(\ref{dyson1}) and (\ref{dyson2}), but one assumes that {\it the size of
the system is  finite} and large. One also assumes that the
initial time of the dynamics has been sent to $-\infty$ and that
{\sc tti} holds. The corresponding construction relies
on the hypothesis that there exists a strong hierarchical structure of
time scales which all diverge with $N$. Calling $t_x$ these time scales,
where $x$ is an index chosen for instance in $[0,1]$,
the hierarchy means
\be
\lim_{N \to \infty} t_x =\infty, \ \ \ \lim_{N \to \infty} {t_x \over t_y}
=\infty
\ \ \ \mbox{if}  \;\;\; x<y
\; .
\label{time_hierarchy}
\ee
The largest of these time scales thus corresponds to the equilibration time
$t_{\sc erg}$.

This allows to produce a solution to the dynamical equations which exhibits
a non-trivial dynamics within each diverging time scale $t_x$. This dynamics
exhibits {\sc tti}, as assumed from the beginning, but it violates the {\sc
fdt}.
 This astute solution presents several formal similarities with the static
solution
of Parisi (and also with the out of equilibrium
study), starting with the hierarchical structure, which acquires an appealing
interpretation in  terms of diverging time scales. However it suffers from
several
problems. At the level of the results first: the correlation function
$C(\tau)$, at very large times $\tau>t_{\sc erg}$, does not go to the correct
Boltzmann Gibbs equilibrium value, known from the static studies. This is
clearly
inconsistent. A careful study of their derivation reveals two weak points.

First, there is an inconsistency in the
hypotheses. If the initial time has been sent to $-\infty$, for a finite
$N$ system, then it is fully equilibrated, and then the dynamics is necessarily
  {\sc tti}
{\it and} obeys {\sc fdt}.  Otherwise (if the waiting time is much smaller than
$t_{\sc erg}$) there is no reason to assume either {\sc tti} or {\sc fdt}.

Second, this approach uses the dynamical equations derived within the $N \to
\infty$ theory, for a finite system where activated processes take place.
Even though  large $N$ saddle-point approaches {\em can} be used to study
activated processes
that occur for large but finite $N$, there are subtleties related to the
existence of multiple solutions.  The multiplicity of solutions is related to
the fact that the results at times greater than  $t_{\sc erg}$ should be non-self
averaging, as the replica solution shows \cite{Ku}.

Because of these problems, which created many discussions
\cite{houghton,Hornerdyn,Dofeio,Russes},
there have been various attempts in the litterature to try to amend this
solution, while keeping most of its nice mathematical structure. One
possibility, first suggested by Horner \cite{Hornerdyn} and developed
in Ref.[\cite{Dofeio}], is to keep the cooling rate finite. This allows
to send the volume to infinity while keeping a regularization time scale
 which is the inverse cooling rate, which is sent to infinity
in the end. What happens once the cooling procedure is over
is still not clear in this approach.
On the other hand, Horner \cite{Hornercoupling} proposed an alternative
`regularizing' procedure which consists in making the disorder time-dependent.
In this way, aging disappears, the solution is {\sc tti} and manifestly
out of equilibrium.

We believe that the present out of equilibrium dynamical approach, inspired by
the experiments themselves,
is very clear. It is seen to work consistently within the finite time, infinite
$N$ regime (\ref{tN}). The price to pay is that one has to abandon the
postulate of {\sc tti}, and think in the two times plane.
In a vague sense, there is also a regularization time which is involved, namely
the age of the system.

Finally we want to point out that in all the approaches developed
so far to study the spin glass phase at the mean-field level, there
is a hierarchical (ultrametric) structure which is involved. 
This is true in the statics where it is hidden in Parisi's Ansatz.
It is also true in the `dynamics on diverging time scale' approach where
it appears in the hypothesis of strong time hierarchy (\ref{time_hierarchy}).
In the out of equilibrium dynamical approach, the situations is rather more
favourable in that
ultrametricity can be proven with  mild assumptions (see \ref{solution} above).

\subsection{On the links between the static and dynamical approaches.
Phase space geometry}
\label{geometry}

In an infinite system, there is no reason for the out of equilibrium
dynamics to be related to the static, equilibrium picture. Indeed
as we saw, the dynamics refer to finite time scales (when $N\to \infty$),
while the static properties are only recovered in the opposite, non-physical
limit \footnote{The divergence of $t_{\sc erg}$ at large $N$ is known to be
of the type $t_{\sc erg} \sim \exp(N^\alpha)$.}. Yet it is
instructive to compare the results of the two approaches. It allows
to gain some intuition  on the
physical mechanism at the origin of aging. It further
 underlines the
physical difference  between the low-temperature behaviour of
the two classes of mean-field spin-glasses.

In continuous spin-glasses one
gets the natural result that the dynamical transition temperature $T_c$
coincides with the static critical temperature corresponding to the
onset of a non zero $q_{\sc ea}$ and of replica symmetry breaking effects.
Furthermore, the large time values  of one-time intensive quantities  such as
the
 energy density
 $E(\infty)$ in the limit (\ref{tN}) coincide with the
value  $E_{\sc eq}$ found at equilibrium within the static (replica) approach.

In discontinuous spin-glasses the results are more surprising:
 the static transition temperature $T_s$
is lower than the dynamic one\cite{Kithwo1,Hornerdyn}, $T_s<T_c$.
In fact the static thermodynamic quantities
computed within the Boltzmann Gibbs equilibrium are
perfectly analytic in the neighborhood of the dynamical
temperature $T_c$,  and conversely nothing special happens at $T_s$
in the dynamics (\ref{tN}).
Furthermore, throughout the low-temperature phase,  the out
of equilibrium dynamical energy does not
converge to the equilibrium one $E_{\sc eq}$. In fact it never goes below a
`threshold level' \cite{Cuku1}
 $E(\infty) \equiv E_{\sc thres}>E_{\sc eq}$.

\vskip 0.5cm
$\bullet$ The energy landscape
\vskip 0.5cm

A geometric explanation for this strange phenomenon can be found within the
framework of the static mean-field equations of Thouless, Anderson and Palmer
({\sc tap} equations). In the static limit
it is possible to write a
free energy  $F_{\sc tap}(m_1,...,m_N)$  in terms of local `magnetization'
variables which represent the average
value of a spin on a large time-window (but the large $N$ limit has
to be taken first). The minima of this
free energy correspond to various metastable states; it is
known that their weighted sum gives back the correct equilibrium
results \cite{ciranoyoung}.

It is far from obvious in general that the dynamical evolution of a system
can be seen as the relaxation of a point in this free-energy landscape. Yet
it is always possible to compute the dynamical properties, such as the
energy, and to see in what region of the landscape the dynamics takes place.
This actually provides interesting insights for discontinuous spin-glasses.
The most complete discussion is available for the case of the spherical
$p$-spin system. It turns out that the {\sc tap} states can be computed rather
easily in this case. The reason is the absence of chaoticity: A {\sc tap}
solution at temperature $T_1$ can be followed adiabatically when one
changes the temperature to $T_2$, the only change is a global rescaling of the
$(m_1,...,m_N)$. Once the free energies
of the solutions are ordered at a temperature $T_1$, this order
is maintained at all temperatures: there is no crossing of the solutions
in the temperature - free energy plane (see Fig. \ref{figtaprev}).
Non-trivial {\sc tap} solutions exist in a wide range of temperatures,
extending above the static transition temperature $T_s$, and even above the
dynamic one $T_c$.

Comparing this landscape to the results of the dynamics, the present
understanding
of the situation is as follows\cite{Kupavi,Cuku1,BBM_TAP}: above the dynamical temperature $T_c$,
there is a coexistence of a paramagnetic state and a bunch of
non trivial, but isolated, {\sc tap} states. When the dynamics starts from
random initial conditions, the system thermalises within
the paramagnetic state. However if one chooses carefully the
initial conditions, one can let the system thermalize within one of the {\sc tap}
states, which are separate ergodic components \cite{BBM_TAP}.

Between $T_s$ and $T_c$, the paramagnetic state is
 fractured into exponentially (in $N$)
many separate ergodic components, each of which has a higher free-energy
compared to that of the paramagnetic state, by an amount which is exactly
equal to their overall configurational entropy, called the `complexity'
\cite{Kithwo2,Nieuwenhuizen}. Correspondingly, the simplest static approach
(see below for more elaborate ones) does not notice
this subtlety, and still describes the system as a simple paramagnet.
In the dynamical approach, starting from random initial conditions, one finds
that the energy actually remains above the threshold
level which is the energy
of the {\it highest} {\sc tap} state (see Fig. \ref{figtaprev}). Therefore in the
limit (\ref{tN}) the system never reaches the energies where it would get
forever trapped into one of the {\sc tap} states. This is the origin of weak
ergodicity breaking in these models.

In the case of continuous spin-glasses the geometry of
the {\sc tap} landscape looks superficially quite clear:
 while above  the transition temperature $T_s=T_c$ the
 free-energy
 has only one minimum, it develops below $T_c$ exponentially many states, which
matches nicely with the Parisi picture. The identification of the low lying {\sc tap}
minima, and of the pure states in Parisi's construction to the
dynamical ergodic components seems inevitable. The puzzle of how to match this
equilibrium picture with the out of equilibrium dynamics containing infinitely
many time sectors is however completely
open, as we already saw in the previous Section. Surprisingly, many purely
static quantities involving
different states (which are in principle mutually inaccessible) coincide with
their dynamical counterparts. In particular, the large time values of one-time
intensive quantities  such as the
 energy density
 $E(\infty)$ in the limit (\ref{tN}) coincide with the
values found at  equilibrium within the static approach. This means that the
set of two dynamical equations
 (\ref{dyson1}) and  (\ref{dyson2}) contain all the information on the replica
symmetry breaking
solution of these systems at low temperatures (A careful numerical
check of this point can be found in Refs.[\cite{Frme,Frme1,Bacukupa}]). Therefore the
study of
these equations may provide
an alternative route to a rigorous study of the spin glass phase.

\vskip 0.5cm
$\bullet$ The geometrical description of mean-field aging.
\vskip 0.5cm

The models we have been describing can be thought of as the motion of
a point in a high-dimensional phase-space landscape (to the extent that we deal
with soft spins).
For discontinuous models, one finds that, surprisingly, the
effect of temperature on the dynamics is very minor, provided that one stays
within the low temperature phase.
 Indeed, the solution of the equations
 is  regular at $T=0$: the dynamics at zero-temperature  is not very different
from the one  at  finite temperature, $T < T_c$. This at once tells us that
activated
processes are not the main ingredient here.
What is then the origin of aging in mean-field models?

At $T=0$, we simply have gradient descent within the basin of attraction of
some
phase-space minimum, and we can discuss \cite{LK} the essence of the problem
without having
to postulate a `free-energy landscape' with  a dynamical meaning.
Since the  basins are high dimensional objects, a random starting
condition will be in the thermodynamical limit practically on a border between
basins, and  hence will remain
close to this border for all finite times without reaching a  minimum.
Now, the border is itself partitioned into basins of attraction of the stable
points on it, which
are the saddles with one negative eigenvalue. The same argument as before tells
us that, again, they will
not reached in finite times:  the system
remains almost on the `border of a border'. One can now iterate this argument,
invoking  borders within borders and
saddles of higher (but smaller than $O(N)$) indices.

The conclusion is that high dimensional (phase-space) wells take long to fall
into. In this sense, the
 above `endless descent' scenario concerns equally the motion of a mean-field
system
within a basin of attraction and the `fall' of a ferromagnet (in any spatial
dimension) into
 one of its two `wells'. Indeed,  both ordinary coarsening and mean-field aging
are
reminiscent of the phase-space model described \cite{BarratM,Ritort} in Section
2, in which aging is
due to the fact that  downhill directions  are always
present, although in decreasing number as time grows.
What we have described so far  is the  fact that a macroscopically different
phase takes
an infinite time to grow in the absence of a driving field, a fact that when
looked
upon from the phase-space point of view is  recognized to hold {\em even at the
mean-field level}.

At finite temperature one  cannot invoke a simple gradient descent,
 but one can  still use the {\sc tap} free-energy  to understand why the ($N=\infty$)
system  ages and never falls
below the threshold level \cite{Cuku1}. The above discussion can be
reformulated as follows: the
density  of eigenvalues of the matrix of
second derivatives in each {\sc tap} minimum is a shifted semicircle law. The
smallest eigenvalue $\lambda_{\sc min}(F)$ decreases as one considers
minima with higher $F$, and becomes zero precisely
for those with  $F=F_{\sc thres}$. Above the threshold, the spectrum for the saddle
points continues to shift, with now $\lambda_{\sc min}<0$,
so that one encounters saddles with more and more negative eigenvalues. If one
makes a cut of the free energy landscape at
different values of $F$, one obtains disconnected `islands'
around each minimum for $F<F_{\sc thres}$. As one raises $F$ just above $F_{\sc thres}$
one crosses saddles with larger and larger
number of negative eigenvalues. Each time, a separatrix develops and the set of
mutually disconnected  components  becomes more
and more connected. One can then picture the ($N=\infty$)
aging system as falling in more and more
disconnected space, hence moving more
and more slowly -- without ever quite stopping, since there are always
directions where the free-energy decreases (while staying above threshold)
\footnote{This is similar to the
idea of `percolation' in phase space, investigated in
Refs. [\cite{Campbell,NewmanStein}]}.

Again, even at finite temperature this scenario is
rather
distinct from the `trap' picture of Section 2: Because $N$ is infinite
(and the model is fully connected)
 the system never reaches the bottom of a trap
from which it could only escape through
thermal activation. For finite $N$, however, activated processes will begin to
play a crucial
role after a finite amount of time (see Section \ref{Conclusion}).

\vskip 0.5cm
$\bullet$ The `marginal stability' criterion.
\vskip 0.5cm

Finally, we would like to mention the possibility of identifying the dynamical
transition temperature
from purely static computations.
This is very useful since static calculations are generally easier
than dynamical ones, in particular when one deals with discrete spin variables.

In continuous spin-glasses, the clearest mathematical characterization
of the onset of the static transition is to consider two identical
copies of the system with the same disorder, coupled through
a small, but extensive, coupling term of strength proportional to $g$.
For instance in the Edwards-Anderson model, one can compute
the partition function $Z_2$ for two spin systems
$s$ and $\s$, coupled through their local
energy densities \cite{coup_rep}:
\be
H_2=-\sum_{(ij)} J_{ij}( s_i s_j+ \s_i \s_j)- g \sum_{(ij)}
(J_{ij}s_i s_j)(J_{ij}\s_i \s_j)
\label{sk2rep}
\ee
The system is in its low temperature phase when the overlap between
the two copies (defined as $-1/(\beta N) \partial \log(Z_2)/\partial g$)
is discontinuous at $g \to 0$. When $g \to 0^+$ this overlap tends to
the Edwards-Anderson order parameter, while when $g \to 0^-$, it
is the smallest possible overlap between different states.

\begin{figure}
\centerline{\hbox{
\epsfig{figure=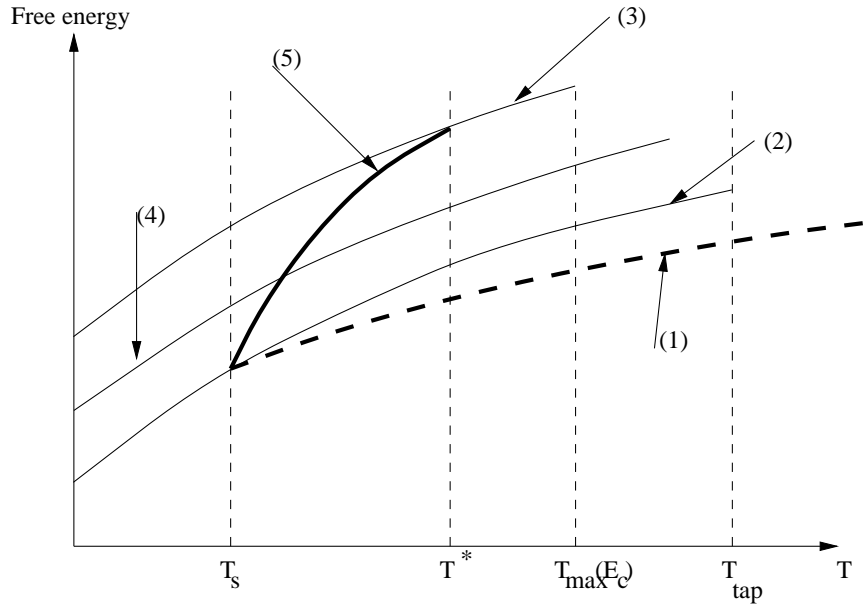,width=8cm,angle=-90}}
}
\caption{A sketch of the free energy of the {\sc tap} states in the spherical p-spin
systems.
Each {\sc tap} state like (2) or (4) can be followed adiabatically in temperature
until it disappears.
The line (1) is the static equilibrium free energy, taking into account the
multiplicity
of the {\sc tap} solutions. (5) shows the free energy of the {\sc tap} states giving the
leading
contribution to the static equilibrium partition function. The highest {\sc tap}
states
(3) are marginally stable and provide the threshold energy: In its
dynamical evolution, the energy of a system starting
from random initial conditions stays above this threshold.}
\label{figtaprev}
\end{figure}

In the case of discontinuous spin-glass transitions this construction
must be modified because of the existence of exponentially
many {\sc tap} states. Roughly speaking the partition function
can be approximated as
\be
Z=\sum_\alpha e^{-\beta F_\alpha}= \int df e^{N(S_c(f)-\beta f)}
\label{ZTAP}
\ee
where $\alpha$ labels various {\sc tap} states, $f=F/N$ are the free energy
densities,
and $\exp(N \ S_c(f))$, usually called the complexity, is the number
of {\sc tap} states at a given free energy density. The large degeneracy of {\sc tap}
states
shifts the dominant region of free energies in the integral (\ref{ZTAP}). If
one took two copies of the system with a small extensive attraction
as in  (\ref{sk2rep}), the leading contribution to $Z_2$ should come
from the case where the two spin systems are in the same state. However, this
leads to $Z_2(g \to 0^+) \simeq
\int df e^{N(S_c(f)-2\beta f)}$
which is dominated by
a wrong saddle point in $f$. Therefore,
in this case, one
must be more clever and study instead $m$ identical copies
of the system in the formal limit $m \to 1$ (in which case, the
correct saddle point in $f$ will be recovered). The onset of a non-ergodic
phase with many components (and thus the value of the dynamical transition
temperature $T_c$), is signalled by the
existence of a non-trivial limit for the order parameter, which is now the
overlap of any two of these $m$ copies in the carefully
ordered limit $\lim_{m \to 1} \lim_{g \to 0^+} \lim_{N \to \infty}$. (In a
replica language what one needs to compute is the free
energy within the `one step rsb' Ansatz, where the $n$ replicas are grouped
into
$n/m$ groups of $m$ replicas, an then expand the resulting free energy at
$n=0$ around the point $m=1$: $ F= F_0+(1-m) F_1 $. One then studies whether
the piece $F_1$ has a non trivial saddle point in the overlap.)

For technical reasons which we shall not explain here, this
criterion is known under the
name of `marginal stability'. It has been used
in many instances for discontinuous spin-glasses. An insightful
interpretation   due to Monasson \cite{Mo} shows how this procedure
amounts to  calculate a partition function restricted to a subset of
configurations, chosen by a random pinning field. A related approach
\cite{Frpa} uses another technique to
 calculate a partition function over a subset of
configurations, where one starts from a typical
 configuration at given  energy called the `pivot',
and calculates the free energy $-1/\beta \log(Z_q)$ associated to all
configurations at overlap $q$ from the `pivot'.
$T_c$ now appears as the temperature where this free-energy  develops a local
minimum for $q \neq 0$.

\section{Glasses and spin-glasses without disorder}
\label{glasseswithoutdisorder}

As emphasized above, the Mode-Coupling equations which have been used with
some success in the recent years to describe supercooled liquids \cite{Go}
formally
coincide with the exact equations describing some mean-field spin-glasses, or
the motion of a point particle in a random potential
\cite{Kithwo,Frhe,Bocukume}
(in large spatial dimension). This is a priori surprising in view of striking
differences between the behaviour and the basic
constitutive ingredients of spin-glasses on the one hand and structural glasses
on the other hand. Let us mention a few rather obvious ones:

-- In spin-glasses, there exists some quenched
disorder (e.g. the random position of the spin carrying atoms in spin-glasses,
 which do not move with time) which is absent in glass forming liquids, where
the disorder is, in a sense
clarified below, `self-induced'.

-- The glass transition is a dynamical effect, basically defined
as the temperature or density at which the relaxation times reaches the
experimentally accessible order of magnitude of hours. The existence of a true
transition at lower temperatures is very controversial. In spin-glasses, on the
other hand, there exists some (experimental \cite{Myd,Nosvrev}
and numerical \cite{Maparu})
evidence in favour of the existence of a second order phase transition (at
least
in zero magnetic field), with a power law divergence of the relaxation time and
of
the non-linear magnetic
susceptibility.

-- Another important
difference is the existence of a crystalline phase in structural glasses, which
can in principle be reached in
a very slow annealing procedure. This has no counterpart in spin-glasses.

In spite of these important differences, there has been
recently an increasing convergence
between the two fields, both at the theoretical level and experimentally,
because of the existence of
an aging regime at low temperatures. This chapter will review some of these
points of convergence.

\subsection{Phenomenology of glasses: a few basic facts}

Many very different glass formers exhibit surprisingly similar properties
when approaching the glass transition. The glass transition temperature
itself is a purely conventional (and somewhat anthropomorphic) temperature
where the relaxation time reaches a value of the order
of $10^3$ seconds. However a common experimental feature is
 the stretching and shouldering
of the relaxation (of, e.g. the density fluctuations) as the temperature is
decreased
\cite{Go,Science} towards $T_g$. More precisely, the
relaxation evolves from a simple Debye exponential at high temperatures
(liquid) to a {\it two-step} process at lower temperature (supercooled liquid),
where the correlation function first decays rather quickly to a
`plateau', and later departs from this plateau value on a much longer time
scale $\tau(T)$. Correspondingly,
the frequency dependent susceptibility $\chi''(\omega)$ evolves from a one
peak, high frequency, structure to a two-peak
structure when the temperature is decreased. The second, low frequency peak
(called the `$\alpha$-peak') shifts to lower
and lower frequencies $\omega_\alpha=1/\tau(T)$ as
$T$ is lowered; the shape of the peak is furthermore strongly non-Debye, which
reflects the fact that the time relaxation functions are non-exponential (and
often fitted by stretched exponentials). The shape of the {\it minimum} lying
in-between these two peaks has been the focus
of an intense interest recently, essentially because one of the
major predictions of {\sc mct} is that, around a certain temperature $T_c$,
\bea
\nonumber
\frac{\chi''(\omega)}{\chi''(\omega_{\min})}
\propto
&
\left\{
\begin{array}{rcl}
\left(\frac{\omega_{\min}}{\omega} \right)^b &\qquad& \omega \ll \omega_{\min}
\; ,
\\
\left(\frac{\omega}{\omega_{\min}} \right)^a &\qquad& \omega \gg \omega_{\min}
\; ,
\end{array}
\right.
\eea
where $a$ and $b$ are two (positive) exponents related through Eq.
(\ref{mctabHT}).
This behaviour reflects, in frequency space, the behaviour of the correlation
function $C(\tau)$ represented in Fig. \ref{revfigChighT2} (see Eq.
(\ref{HighTmct})).

The relaxation time  $\tau(T)$ grows
extremely fast as the temperature is decreased,
 in general faster than the Arrhenius law $\exp(\Delta/T)$. Systems for
which $\tau(T)$ are close to an Arrhenius behaviour are called `strong',
 whereas systems for which the divergence is faster
are called `fragile' \cite{Science}. For the latter systems, a widely used
description of
the experimental data (based on dielectric measurements, viscosity
measurements, etc.) is the Vogel-Fulcher law
\be
\tau(T) \sim t_o e^{\Delta \over T - T_0}
\ee
which suggests that $\tau(T)$ actually diverges when $T \to T_0$, i.e. that
there is a true phase transition at $T=T_0$. This temperature furthermore
appears to coincide with the temperature
(called Kauzman temperature) at which the extrapolated
 excess entropy (as compared to the crystal) would vanish. However, the
divergence of $\tau(T)$ is
so rapid that $\tau(T)$ becomes larger than experimental time scales
 at the glass temperature $T_g$ which is often
appreciably larger than $T_0$. Is is thus difficult to claim the existence
of a transition on the basis of the fit of $\tau(T)$ only (see however
\cite{Nagel}). In particular, other
functional forms, such as $\tau(T) \sim
t_o \exp((\Delta/T)^2)$, also give reasonable fits of the data \cite{Bass} --
without invoking the existence of a critical temperature where $\tau(T)$
would diverge \cite{Cummins}.

\subsection{Discontinuous spin-glasses: a mean-field scenario for structural
glasses}

In order to understand better why some theoretical ideas emerging
from spin-glass mean-field theory might also be relevant for
structural glasses, it is important to keep in mind the fact that, as
we already emphasized in Section \ref{sectionmeanfield}, there exist two
different classes of
mean-field  spin-glasses, continuous and discontinuous. While the most
conventional -- continuous --
ones provide a good starting point for the description of
real spin-glasses, with a second-order phase transition where the
Edwards-Anderson order
parameter vanishes, some discontinuous models are characterized by a {\it
discontinuous} static phase
transition at a temperature $T_s$, where the order parameter is
finite just below $T_s$. An extreme example of this type of spin-glass ordering
is provided by the Random
Energy model \cite{BREM},
 which has zero entropy density in the
whole low temperature phase. There are in fact many other such examples
(see Section (\ref{AvsB})). As discussed there, the discontinuous models
generally
possess a rather peculiar dynamical behaviour, with a dynamical transition
temperature $T_c$ which is {\it higher} than the static one (and which
coincides with the Mode-Coupling critical
temperature). When approaching $T_c$ from above, the relaxation time $\tau(T)$
diverges as a power law, but there is no
singularity in the static thermodynamic quantities, which
are analytic around $T_c$ (see Fig. \ref{CritTemp}).
Thermodynamic singularities, including a jump in the specific heat,
only occur at the lower temperature, $T_s$.  As explained above (Section 3.8),
this behaviour originates from the
fact that as soon as $T < T_c$, the system starts aging for ever in a slow
descent towards a state with a free-energy
extensively higher than the one of the equilibrium state. Note that between
$T_s$ and $T_c$, the number of
such metastable states
is exponentially large, the associated configurational entropy (the
`complexity') being exactly equal to the free energy difference
between the metastable states and the
equilibrium (paramagnetic) state.

\begin{figure}
\centerline{\epsfxsize=10cm
\epsffile{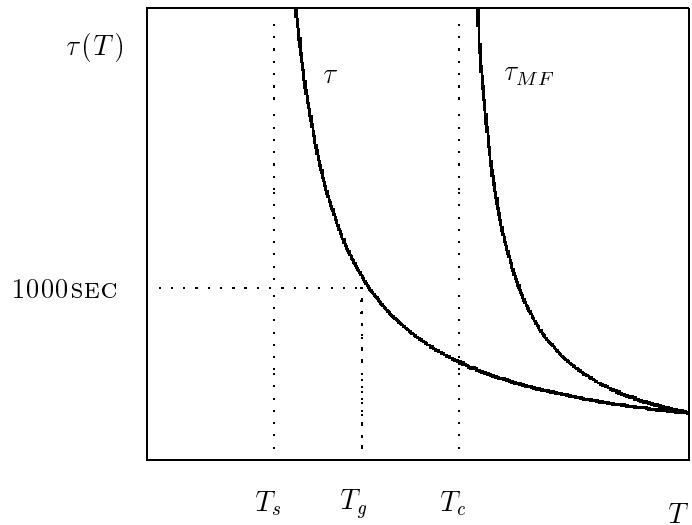}
}
\caption{Relaxation time {\it vs} temperature in discontinuous spin-glasses.
 The right hand curve is the mean-field
prediction, which gives a dynamical transition at a temperature $T_c$ above
the static transition temperature $T_s$. The left curve is a conjecture on the
behaviour
in finite dimensional systems: activated processes smear the dynamic
transition. The
relaxation time diverges only at the static temperature $T_s$, but becomes
experimentally
large already around the glass temperature $T_g$. }
\label{CritTemp}
\end{figure}

Such a scenario can however only exist, strictly speaking, at the mean-field
level, where nucleation barriers are infinite.
In finite dimensional space, metastable states with a free energy density
larger than that of
the ground state have a finite lifetime, since the nucleation of large
`bubbles' of the ground state
are always favoured. The gain in free energy for
a bubble of size $L$ scales as $L^D$, while the surface energy
cost scales at most like $L^{D-1}$ (if the interaction is short-ranged).
Therefore the mean-field
picture cannot survive as such in a finite dimensional system. However, one can
argue that around
the mean-field $T_c$ the
bubble nucleation will be a slow process, leading to a rapid increase of the
terminal relaxation time $\tau(T)$, which only diverges when the static
transition temperature $T_s$
is reached. This picture was proposed nearly a decade ago in an insightful
series of papers by Kirkpatrick,
Thirumalai and Wolynes \cite{Kithwo}$^-$\cite{Kithwo2}: within this hypothesis
$T_c$ is a crossover
temperature below which activated processes become important
\footnote{Similar ideas can be found within the context of `extended' {\sc
mct}, \cite{Go} although the
precise relation between the two pictures is not obvious to us.} leading to a
rapidly
increasing ({\it \`a la} Voger-Fulcher \cite{ParisiVF}) relaxation time as the
temperature is further reduced. The static
ordering critical temperature $T_s$ appears as a Vogel-Fulcher or
Kauzman temperature (below which freezing is complete), while the experimental
glass temperature $T_g$,
where $\tau(T)$ reaches $10^3$ seconds, lies
somewhere between $T_s$ and $T_c$ (see Fig.\ref{CritTemp}).

This rather appealing idea however suffers from important theoretical
loopholes.
First of all, it relies on a model with quenched disorder, absent in structural
glasses.
As discussed in the next paragraph, this might not be too serious as
this disorder might well be `self-induced' by the system itself. One would
actually like to be sure that the above nucleation arguments are indeed correct
for finite dimensional versions of  discontinuous models (there exist
a few numerical simulations for the
Potts glass, $p$-spin spin-glass and `frustrated percolation' models
\cite{Carmesin}$^-$\cite{Coniglio}). The subtlety comes from the fact
that the nucleation is in the present case rather peculiar, since the
nucleating phase cannot
be the ground state (otherwise the system would be completely frozen after a
finite time,
and would loose the contribution of the complexity to the free-energy),
but rather {\it another} metastable phase with
exactly the same free-energy density -- which makes it hard to understand why
the bubbles
should grow at all. The meaning of the `entropic driving force' invoked in Ref.[\cite{Kithwo2}]
is not very clear to us, and, surprisingly, little progress has been made to
support this conjecture \cite{ParisiVF}.
Another picture, somehow related to that of Ref.[\cite{Tarjus}], is that the
complexity induces a
microphase separation into `grains',
each of a different metastable phase, with a certain temperature
dependent size. The relaxation time would correspond to the time needed to a
`grain' to disappear or for a new grain to appear. In this sense, glass
dynamics might have a lot in common with foams or microemulsions,
as recently advocated in Ref.[\cite{Cates}].

\subsection{Self-Induced Quenched Disorder: Spin glasses without disorder}

An important obstacle if one wants to convert the above picture
valid for some spin-glasses into a theory for structural glasses is the meaning
of the quenched disorder in the latter case. It turns out however that a series
of recent works
\cite{JPBMM}$^-$\cite{rods}
has shown the existence of discontinuous spin-glass like behaviour in systems
with frustration but {\it without quenched disorder}.
These systems thus provide natural spin analogues of
glass formers. Although their microscopic description does remain remote
from that of structural glasses (in particular because they involve infinite
range interactions), they
provide at least an existence proof to the phenomenon of self-induced disorder,
and their study is worth the effort, both for their intrinsic
beauty and as a source of inspiration for modelling structural glasses.
Furthermore, from an experimental point of view, Charge Density Wave systems
(among others \cite{Chandra})
have recently been shown to behave
very much like disordered systems \cite{Biljakovic}, with however a very small
density of impurities, suggesting
that incommensurability effects alone (inducing some frustration) might be
sufficient to generate `self-induced disorder' \cite{Aubry}.

\begin{figure}
\centerline{\hbox{
\epsfig{figure=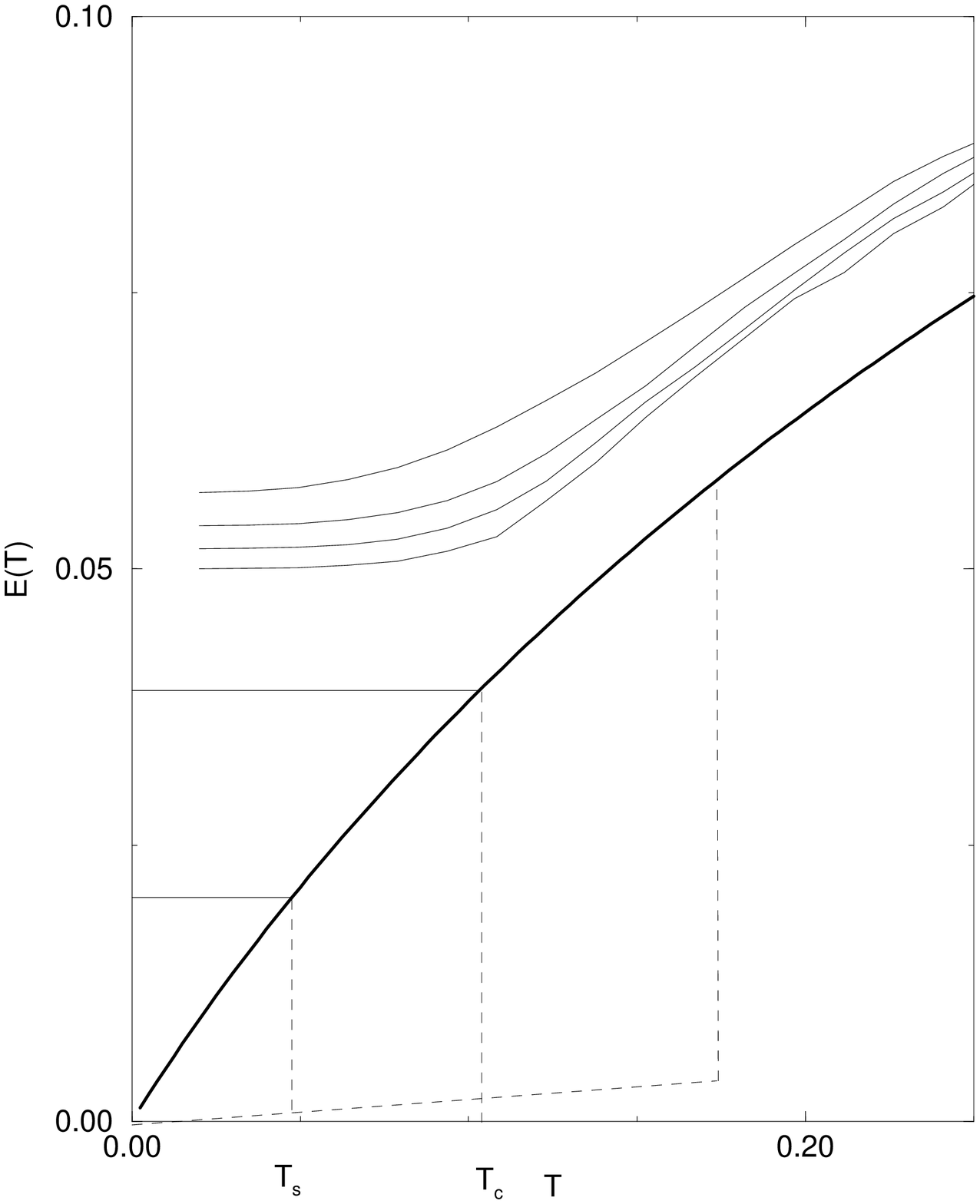,width=6cm}}
}
\caption{The energy of the {\sc labs} model with open boundary conditions vs
temperature.
The top curves are the results from a Monte-Carlo
simulation of a $N=401$ spin system  with logartihmically decreasing cooling
rates (from Ref.[158]).
The other curves are derived analytically with the fiduciary random system.
$T_c$ is the
dynamical transition temperature below which the fiduciary system freezes, and
$T_s$ is
the static transition temperature which cannot be found from a Monte Carlo, but
is accessible
in principle from exact enumerations of systems with a much smaller size. The
lowest curve,
corresponding to the 'crystalline' state,
has been found only in the case of periodic boundary conditions and for some
special values of $N$.
}
\label{energieLABS}
\end{figure}

Monte Carlo simulations show that the low-temperature dynamics is highly
non-trivial
in all cases.
Although  an analytical study  of the dynamical features could in principle be
done
along the lines described in Section \ref{sectionmeanfield}, in many of the
recent papers the shortcut was taken
of calculating some dynamical properties using the `pseudo-statical' approaches
we described in Section \ref{statics}.

\vskip 0.5cm
$\bullet$ Low autocorrelation binary sequences.
\vskip 0.5cm

We shall first concentrate on some spin systems with frustration
but without disorder, which contain long-range interactions. These
systems exhibit the
same behaviour as the discontinuous mean-field spin-glasses, namely
a dynamical transition at a temperature $T_c$ larger than the
temperature $T_s$ of the static transition. The first example is the problem of
`low  autocorrelation binary
sequences' ({\sc labs}). This is an old and important problem from
communication theory
\cite{golay} which was
restated in physical terms by Bernasconi \cite{bernas} as follows:
take a one dimensional chain of Ising spins $\s_i=\pm 1, \ i=1...N$. Compute
the
correlation function at distance $k$
\be
C_k=\sum_{i,j=1}^N \s_i \s_{j} \delta_{j,i+k}
\ee
and define the energy function as
\be
E(\{\s\})={1 \over 2(N-1)} \ \sum_{k=1}^{N-1} C_k^2
\; .
\ee
 The interest in communication theory
is to find the low energy configurations. In fact the ground state of this
energy function does provide a sequence of bits which minimizes
 the two point correlations, and this is also useful for building a
pseudo random number generator. Two versions of this problem
have actually been studied, differing in the choice of boundary
conditions. Due to the infinite range of the interactions, they present
significative differences. A first version studied in Refs.[\cite{JPBMM,mpr1,kramez}]
has free boundary conditions,
where the correlation function $C_k$ is defined with the sum over
the spin indices $i$ going from $1$ to $N-k$. Another choice, studied in
\cite{mpr1} is that of periodic boundary conditions, where one
can define $C_k$ through a sum over
the spin indices $i$ going from $1$ to $N-1$.  As any other optimisation
problem, this can be generalised to a finite
temperature study by assigning to each sequence of spins a Boltzmann weight
$P(\{\s\})=\exp(-\beta E(\{\s\})/Z$.

Let us first discuss the case with free boundary conditions.
 It was shown by Monte Carlo simulations that
there exists a finite temperature freezing region in
a temperature range around $T \simeq 0.1$, with a weak cooling rate
dependance of the low temperature energies (see Fig. \ref{energieLABS}).
Computations spanning
very long time scales have been perfomed at low temperature by using an
efficient Monte Carlo
algorithm \cite{kramez}$^-$\cite{Bortz}, and reveal a
clear aging effect, characterized by a $\tau/t_w$ scaling (see Fig.
\ref{agingLABS}).
The smearing of the transition and the cooling rate dependence might be
a finite $N$ effect.
In this  case, the presence of `traps' in
phase space with a broad distribution of trapping times \cite{kramez}
is rather convincingly observed for finite
$N$ (see Fig.\ref{trapsLABS}).

\vskip 0.5cm
$\bullet$ A replica analysis for non-disordered problems.
\vskip 0.5cm

The analytical study of the {\sc labs} model is in itself very interesting.
Despite its simplicity, we know of no
direct solution. A rather indirect, but illuminating, way of proceeding is to
replace the non-disordered {\sc labs}
model  by
a  `fiduciary' model with quenched disorder. The basic idea consists in
considering the model at hand
as one special sample of an ensemble of systems containining quenched disorder.
In the case with free boundary
conditions this is achieved as
follows \cite{JPBMM,mpr1}: one defines a `disordered' correlation function
\be
C_k^d= \sum_{ij} {M^{(k)}}_{ij} \s_i \s_j \ \label{disorderedlabs},
\ee
where $M^{(k)}$ is
a matrix with random elements, equal to $0$ or $1$, with the only
constraint that $\sum_{ij} M^{(k)}_{ij} =N-k$. The original problem is
a particular choice of $M^{(k)}$, where the only nonzero
elements are on the $k^{th}$ diagonal. The hope is that this particular case is
a generic case,
and this is actually not at
all obvious. (For example, it would be nonsense to claim that a ferromagnet is
a special instance of a spin-glass
with $J_{ij}=\pm J$ couplings, where all
$J_{ij}$ happen to be equal to $+J$: the ferromagnet is
just a very atypical sample.)
 There is in fact quite a bit of educated guesswork involved in the choice of
the
ensemble of disordered system, of which the original model is argued to be a
generic member -- see below. In
the present case, the original model is extremely frustrated due to the
long-range and conflicting nature of the
interactions, two features which are indeed retained
by the Hamiltonian defined using Eq. (\ref{disorderedlabs}).

Now, the crucial remark is that if the model is indeed generic, its static
properties   can be obtained
by means of the  replica method, where the averaging is performed over the
fictitious disorder. In the case at hand, the resulting free energy indeed
turns
out to be a good approximation of the original model in the high temperature,
replica symmetric phase. This approximation actually corresponds to the one
proposed by
Golay \cite{golay} using different arguments; as seen from a
high temperature expansion, this approximation is however not exact (but see
next paragraph). Its main virtue
is to predict the existence of a static phase transition at a temperature
$T_s=0.0476$, below which a breaking of replica symmetry of the discontinuous
type
appears (see Fig. \ref{energieLABS}).
The low temperature phase is characterised
by a residual entropy density which is linear in $T$, but small (less than
$10^{-5}$ per spin
at $T_s$). From a glass point of view this phase transition
can be seen as the resolution of an entropy crisis appearing
at an extrapolated Kauzman temperature which is very close to $T_s$. The
prediction
for the ground state energy density, $E_0/N \simeq 0.0202$, is compatible with a
large $N$ extrapolation of the ground state
energies found by exact
enumeration on small samples \cite{mertens,bernas} $N \leq 48$. On the other
hand,  it does substantially differ from the
apparent ground state
energy extracted from Monte Carlo simulations, even after extrapolating to
very small cooling rate (see Fig.\ref{energieLABS}). In fact, similar
discrepancies have been seen and studied in detail before on several
disordered spin-glasses (for instance, the binary perceptron problem
\cite{hornerperc}). This is again reminiscent of the above
discussion of discontinuous spin-glasses and of the existence of a dynamical
transition at $T_c>T_s$, where the
Langevin (or Monte Carlo) dynamics gets trapped in metastable
states with high energies $E_{\sc thres}=E(T_c)$. In analogy with other
discontinuous spin-glasses, one may thus expect that the dynamical
transition where the energy freezes takes place at a temperature $T_c$ fixed by
the marginality condition (see Section \ref{statics}), i.e the temperature
where
a replica symmetry broken solution first appears, with a breakpoint  (in
Parisi's Ansatz) equal to  $m=1$. This leads to \cite{JPBMMunpub}
$T_c=0.103$, in reasonable agreement with the Monte-Carlo data (see Fig.
\ref{energieLABS}).
\vskip 0.5cm

\begin{figure}
\centerline{
\epsfxsize=8cm
\epsffile{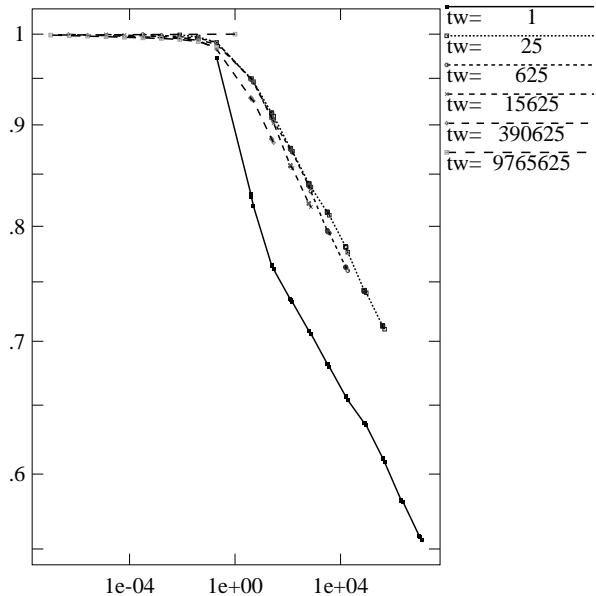}
}
\caption{Two time correlation $C(t_w+\tau,t_w)$ obtained in an extensive Monte
Carlo simulation
of the {\sc labs} model of $400$ spins with open boundary conditions
(Ref. [158]),
at a temperature $T=.075$, plotted versus $\tau/t_w$. The data exhibits a clear
aging effect.
}
\label{agingLABS}
\end{figure}

\begin{figure}
\centerline{\hbox{
\epsfig{figure=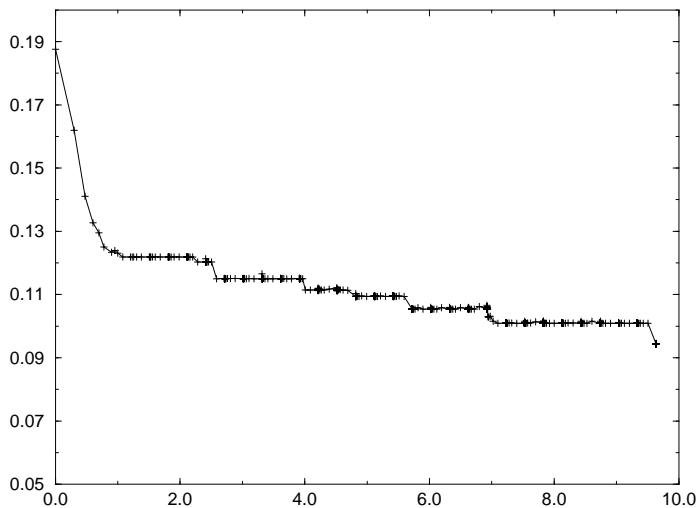,width=8cm,angle=-90}}
}
\caption{Energy per spin {\it vs} $\log_{10}$ (time)
 in a single Monte Carlo run for a $N=400$  {\sc labs} spin system at $T=.075$
(From Ref.[158]).
}
\label{trapsLABS}
\end{figure}

$\bullet$ A model with a `crystalline' state.
\vskip 0.5cm

In summary, we have shown that the {\sc labs} model with free
boundary conditions provides an interesting example of a non disordered
mean-field spin system, sharing many similarities with discontinuous spin
glasses. However, the fiduciary
Hamiltonian constructed using the disordered correlation Eq.
(\ref{disorderedlabs}) is only
 approximate, and does not, for example, give the exact free-energy in the high
temperature phase.
It turns out that the situation is under better control with {\it periodic}
boundary
conditions \cite{mpr1,mpr2}. In this case, the model furthermore exhibits,
in some sense, a `crystal' phase.

The simplification comes from the fact that with periodic boundary conditions,
the energy can be expressed as
\be
E = {1\over N^2} \sum_{q=1}^{N} |s(q)|^4 \ ,
\label{E_per}
\ee
where we have introduced the Fourier series
$\s(q)=\sum_j \s_j \exp(2i\pi qj /N)$. The problem remains non trivial since
one has to remember that $s(q)$ is constrained to be the Fourier transform of a
{\it binary} sequence.
In this case a clever choice of the fiduciary disordered system was found in
\cite{mpr1}, which consists in
substituting the usual Fourier components $\s(q)$ by a disordered version of
the Fourier transform:
\be
\s^d(q)=\sum_j U_{q,j} \s_j
\ee
where $U$ is a kind of random unitary matrix
\footnote{
There is a subtlety in this construction, namely the
fact that one wishes to introduce a 'disordered Fourier
transform' with a matrix $U$ which is unitary, but also
satisfies $U_{-q,j}=U_{q,j}^*$, where $^*$ stands
for complex conjugation. The proper construction through
random orthogonal transformation is described in Ref.[\cite{mpr1}].
}. One can again compute the equilibrium properties of
the disordered problem with the
replica method. One finds, similarly to
the case with free boundary
conditions, two transition temperatures $T_s$ and $T_c$. However, the
periodic {\sc labs} problem is in fact richer \cite{mpr1,migrit}:

--  It can be shown that the disordered model has the same free energy as the
original one
{\it to all orders} in a high temperature expansion. The phase transitions,
both dynamical and static,
predicted by the replica solution of the disordered problem are
in good agreement with the numerical simulations. The identity
of the two problems in the low temperature phase, at the level
of extensive thermodynamic quantitites, is however still a conjecture.

-- It turns out that the periodic {\sc labs} model possesses a very special
ground state for some
particular values of $N$. When $N$ is prime, clever number theoretic properties
can be used to generate a sequence
of spins with a finite energy $E$, and therefore a vanishingly small energy
density $E/N$ in the thermodynamic limit. (Actually for $N$ prime of
the form $4k+3$, $k$ integer, the ground state energy  is $E=1$, and for $N$
prime of the form
$N=4k+1$, is is found to be $E=5$). These special configurations are very
difficult to find using
Monte-Carlo like dynamics: the energy landscape has sometimes been described as
a `golf course'
potential, in the sense that the ground state is an unexpected
deep hole surrounded by rather unfavorable states. These ground states
constitute the analogue
of a crystalline state. If one simulates a {\sc labs} model with a `magic $N$'
(in the above sense),
starting at zero temperature from
the crystalline state, one finds that the low temperature specific heat is very
small, and the energy
has a discontinuous jump (first order transition) to the high temperature
energy curve at a `melting'
temperature $T_m>T_c$ (a sketch is given in Fig. \ref{energieLABS}). No such
crystalline state
has been found in the free boundary case \cite{mertens}.

\vskip 0.5cm

$\bullet$ Speculations on the `fiduciary' disordered Hamiltonian strategy.
\vskip 0.5cm

We have thus seen how some frustrated spin systems without
disorder can be solved (approximately, or even exactly at least in the high $T$
phase
 for the {\sc labs} with periodic boundary conditions), following
a rather interesting strategy. This strategy consists in
substituting the original problem by a `fiduciary' one with quenched disorder,
and solving the disordered system using, e.g., the replica method to obtain the
static properties
and information about the transitions.
There is
unfortunately no systematic method of
choosing the fiduciary model so far. The above two examples, or other models
which have been studied in a similar way
\cite{mpr2,Cukumopa}, show
the importance of symmetry considerations in the choice of the fiduciary
disordered problem,
and suggest as a criterion that this disordered model should be as `close' as
possible to the
original one in the high temperature (liquid) phase. This strategy is
reminiscent of the
very fruitful approach to energy levels in complex nuclei
through the study of fiduciary
random Hamiltonians with the proper
symmetries \cite{dyson}. In our case we do not
yet understand when such an approach may be successful or not, if it is only
restricted to finite time dynamics or if it does apply to thermodynamical
properties.
In some cases (see \cite{Frhe,mpr2,joseph} and below), the `spin-glasses
without disorder' explicitly involve pseudo random numbers in the sense that
the spin
couplings are deterministic, but very rapidly oscillating. This is much less
obvious in the
{\sc labs} model, especially with free boundaries. Finally, as discussed in the
next subsection,
self-consistent (Mode-Coupling) approximations of non disordered models often
lead to
equations which are exact for some adequately chosen disordered systems.

In a loose sense, one expects that the slow dynamics at low temperatures
originates from some degrees of freedom which freeze and
 play the role of an effectively quenched disorder field for the other degree
of freedom.
 The success of the present strategy might lie in the fact that the precise
identification of
these `slow' variables is not necessary to understand the
freezing transition in these models.

\vskip 0.5cm
$\bullet$ Other mean-field spin-glass models without disorder.
\vskip 0.5cm

There exist by now a few such examples of spin-glasses without disorder.
Besides the `fiduciary' Hamiltonian strategy, other approaches have
been attempted, in particular some direct diagrammatic calculations -- either
static ones \cite{Cukupari,joseph} (basically through a resummation
of the high temperature expansion) or using dynamical
perturbation theory \cite{Frhe,joseph2}. Because of space limitation
we cannot mention them all here. We should nevertheless point out
 the case of Josephson junction arrays because it may allow a direct
experimental investigation of many of the ideas
discussed here \cite{joseph,joseph2}. This model involves two sets
of $N$ spins, $\s_j=\exp(i \phi_j)$ (living on the rows of a 2-D array of
Josephson junctions) and $\s'_k=\exp(i \phi'_k)$
(living on the columns of the
same array) which are normalised two component
vectors (U(1) spins). These are coupled through the Hamiltonian
\be
H=-{J \over \sqrt{N}}\sum_{j,k=1}^N
\(( \exp(i 2 \pi \alpha j k/ N) \s_j^* \s'_k + c.c. \))
\ee
where c.c. means complex conjugate. This can be realised by using a
stack of two mutually perpendicular sets of $N$ parallel
wires. There is supposed to be a Josephson junction
at each crossing of an horizontal wire with a vertical one.
The variable $\phi_j$ is the reference phase of the $j^{th}$ horizontal
wire, while $\phi'_k$ is the reference phase of the $k^{th}$
vertical wire. The system lies in an external transverse field $H$,
which induces a phase shift per unit area $\alpha$ proportional to $H$.
The high temperature expansion has been resummed for $1/N \ll \alpha \ll 1$,
both for the static theory \cite{joseph} and for the
two-time correlation and response function \cite{joseph2}, leading
{\it exactly}
to the same equations as those
describing the $p=4$ spherical spin-glass which we have discussed in Section
\ref{sectionmeanfield}.
Again, the system undergoes two phase transitions, a dynamical one at a
temperature $T_c$ larger than the static one $T_s$.
These dynamical equations are also equivalent
to a particular case of the `schematic' mode coupling equations of supercooled
liquids. Josephson junction arrays may
thus provide an interesting experimental playground to test directly the
predictions of these theoretical studies.

\vskip 0.5cm
$\bullet$ Towards the description of the glass phase for interacting particles
\vskip 0.5cm

As is often the case in statistical physics, it is technically easier to study
a phenomenon (here the glass transition in spin  systems without  disorder)
on the magnetic case. A natural problem is to try to extend the kind of ideas
that we have seen at work on these spin analogues towards more realistic
problems where a glass phase appears without any quenched disorder, like
glass forming liquids.

One must thus describe interacting particles. The studies in this direction are
still
rather preliminary. Two routes have been recently explored.
One is to consider interacting particles in large dimensional ($D$) spaces.
In Ref.[\cite{Cukupari}]  $n$ point-like particles interacting repulsively
are constrained to move on the vertices of a $D$-dimensional
hypercube. The model can be mapped onto a modified  $O(n)$ matrix model.  It
has a sharp glassy
transition of the kind described previously.
Amusingly, for $n$ multiple of four, an unproven  conjecture of Hadamard
gives the `crystalline' ground states.
On the same line, similar properties are found in a model \cite{Cukumopa} of
hard spheres moving on the surface of a $D$-dimensional hypersphere.

Another study retains the three dimensional nature of the problem, and starts
from a reasonable analytic approximation
of the liquid phase given by
 the Hypernetted Chain Approximation ({\sc hnc}). It turns out that one can
generalize this
{\sc hnc} description through the introduction of a replicated theory as
described in
Section \ref{statics}. This approach \cite{hnc} yields a good analytical
prediction for
the glass transition temperature or density of hard or soft spheres systems.
A precise description of the low temperature phase along the same lines is
however still missing.

\subsection{p-spin models, Mode Coupling Theory of glasses, and its extension
at low temperatures}
\label{modecouplingglass}

The above arguments  suggest that discontinuous mean-field spin-glasses,
 in spite of the presence of quenched
disorder, can provide a good starting mean-field theory for structural glasses.
Among the most striking convergence
between the two subjects, already alluded to many times above, is the
equivalence (in the high temperature phase)
between the Mode Coupling equations for glasses and the dynamical mean-field
equations for $p$-spin-glass models. This analogy
was first noticed already long ago \cite{Kithwo}. It has been useful
technically, in particular by transposing the mathematical analysis of the Mode
Coupling equations developed for glasses \cite{Go,leut}  to the study of
the high temperature dynamics of  $p$-spin systems \cite{Crhoso} and manifolds
in random media \cite{Kiho1,Kiho2}, e.g.
the divergence of the relaxation time and the shouldering of the
relaxation. Following similar developments in the field of fully developed
turbulence
\cite{kraich1,kraich2}, it has been realized more recently that the
factorization property which
is at the heart of the mode coupling approximation actually becomes {\it exact}
for certain systems, which turn out either
to contain quenched disorder \cite{Bocukume} or some deterministic version of
disordered systems, in the sense of having rapidly oscillating  couplings which
have statistical properties of
disordered ones \cite{Frhe}.
We shall first give a general flavour of why this is the case,
and then turn to the implications of this general result to the low temperature
extension of the Mode-Coupling equations, and its physical consequences.

\vskip 0.5cm
$\bullet$ Mode coupling approximation and hidden disorder.
\vskip 0.5cm

Both glass forming systems and turbulence can be described
by some non-linear stochastic dynamical equation. In order to
describe the essence of the mode coupling approximation and its relation
with disordered systems, we shall explain it briefly
on
the simple case of a single scalar degree of freedom $\phi$,
with a Langevin dynamics
\be
{\partial \phi \over \partial t} = - \mu(t) \phi - {g \over 3!} \, \phi^3 +\eta
\label{2}
\ee
with initial condition $\phi(t=0)=0$. The thermal noise $\eta$ is a Gaussian
noise $\eta$ with $\langle \, \eta(t)  \, \rangle = 0 $
and $\langle \, \eta(t) \,
 \eta(t') \, \rangle =
2T \, \delta(t-t')$ (in the following the brackets will always denote
an average over the realisations of the Gaussian white noise $\eta$).
The coupling constant $g$
serves as a book-keeping parameter to set up a perturbative expansion. This
expansion
can either be well-behaved or  ill-behaved depending -- say -- on the dimension
of
space. It is in any case rather useless when $g = O(1)$ if it cannot be
resummed in one
way or another. The mode coupling
consists in a `one-loop'
self-consistent perturbation theory.
This amounts to resumming a particular (infinite) set of terms
in the perturbation expansion. In this way, non-trivial self-consistent
equations are
obtained, which enable one to peep into the strong coupling regime, through an
approximation which is however not easily controlled.

Setting
$R_0 = [\mu(t) + {\partial  \over \partial t}]^{-1}$, which gives
$R_0(t,t')=
 \exp \left( -\int_{t'}^t d\tau \; \mu(\tau) \right) $,
 the perturbative expansion for $\phi(t)$
is easily written as:
\be
  \phi(t) =
R_0 \otimes \eta
-
\frac{g}{3!}  \; R_0 \otimes \{ R_0 \otimes \eta \bullet R_0 \otimes \eta
\bullet R_0
\otimes \eta   \} + ...
\label{3}
\ee
where $\otimes$ means a time convolution: $(R_0 \otimes f)(t) = \int_0^t dt'
R_0(t,t') f(t')$ and $\bullet$ is a simple product.

The lowest non trivial order perturbative expansion for the
correlation function
 $C(t,t')$ and the response
function $R(t,t')$ is easily written in terms of
the kernels $\Sigma(t,t')$ and $D(t,t')$ through the Dyson equations:
\beqa
R(t,t')
&\equiv&
 R_0(t,t')+ \int_{t'}^t dt_1 \int_{t'}^{t_1}  dt_2 \; R_0(t,t_1) \;
\Sigma(t_1,t_2) \; R(t_2,t')
\; ,
\label{DysonG}
\\
C(t,t')
&\equiv&
  \int_0^t dt_1 \int_0^{t'}  dt_2 \; R(t,t_1) \; D(t_1,t_2) \;
R(t',t_2)
\label{DysonC}
\; .
\eeqa

The mode coupling approximation for this problem amounts to an approximation of
the kernels
 $\Sigma(t,t')$ and $D(t,t')$ where one takes their
values at order $g^2$ and substitutes in them the bare propagator
$R_0$ and the bare correlation by their renormalised values. This
gives the following self-consistent equations:
\beqa
\Sigma(t,t')&=& {g^2 \over 2} \; C^2(t,t') \, R(t,t') \nonumber \\
D(t,t')&=&2T \; \delta(t-t')+ { g^2 \over 6} \; [C(t,t')]^3 \ \ ,
\label{11}
\eeqa
 This approximation
 neglects
 `vertex renormalisation'.

 The problem is of
course to try to control this procedure. An important step in this direction is
to
identify a model for which the self-consistent equations are {\it exact}.
The basic remark (first made by
Kraichnan in the context of turbulence \cite{kraich2}, where the analogous
method is named direct interaction approximation)
is that the diagrams retained by the mode coupling approximation are precisely
those which
survive if one considers the following disordered problem. First, one
upgrades $\phi$ to an $N-$ `colour' object $\phi_\a$, where $\a=\{1,2,...,N\}$.
The equation of motion Eq. (\ref{2}) is then generalized to:
\be
{\partial \phi_\a \over \partial t} = - \mu(t) \,  \phi_\a
-
4 g \; \sum_{\b < \g < \delta} \jabgd \;
\phi_b \phi_\g \phi_\delta +\eta_\a
\label{7}
\ee
with independent noises $\eta_\a$.
The couplings $\jabgd$ are {\it quenched}, i.e. time-independent,
Gaussian random variables of zero mean and variance  $\overline
{J_{\a\b\g\delta}^2} =  1/N^3$.
In the large $N$ limit,
the correlation:
$
C(t,t')
\equiv
\frac{1}{N} \;
\sum_{\alpha=1}^N \; \overline{\langle \phi_\a(t) \phi_\a(t') \rangle}
$
 (where the overline denotes the
average over
the random couplings $\jabgd$)
and the response:
$
R(t,t')
\equiv
\frac{1}{N} \;
\sum_{\alpha=1}^N \;
\overline{\langle {\partial \phi_\a(t)\over \partial \eta_\a(t')} \rangle}|_{\eta=0}
$
precisely obey the mode coupling approximation equations \cite{Bocukume}, Eqs.
(\ref{DysonG})  and (\ref{DysonC}).
\vskip 0.5cm
$\bullet$ A particle in a random potential.
\vskip 0.5cm

The same construction can be generalised \cite{Bocukume} to an arbitrary
nonlinearity $F(\phi)$ substituting the ${g \over 3!} \, \phi^3$
in Eq.(\ref{2}). Therefore one finds that the
general schematic mode coupling equations developed in the study
of glass forming liquids can be derived exactly from the Langevin dynamics of
 $N$ continuous spins
$\phi_\a$, of the
type:
\be
{\partial \phi_\a \over \partial t} = - \mu(t) \,  \phi_\a
-\frac{\delta V[\{ \phi \}]}{\delta \phi_\alpha}
+\eta_\a
\; .
\label{12}
\ee
This disordered multispin Hamiltonian is precisely the generic mean-field
problem (\ref{modelito}) which we studied in Sect. \ref{sectionmeanfield},
 which can also be seen
as describing a particle evolving in an $N \to \infty$ dimensional space
in a quenched random potential $V[\{ \phi \}]$.

This interpretation is rather appealing. Let us introduce the following highly
simplified picture of a glass: the motion of a given particle can be thought of
as
taking place in a random potential created by its neighbours. Since the motion
of the molecules is extremely slow at low temperatures, one can assume that this
random potential has a static component, in the spirit of the `self-induced
quenched disorder' scenario
which we discussed above \cite{BCM}. In large dimension of space, one can
establish the {\it exact} equations relating
the two-time correlation function $C(t_w+\tau,t_w)=\langle\vec r(t_w+\tau)
\cdot \vec r(t_w)\rangle$
(where $\vec r(t)$ is the position of the particle at time $t$), and the
two-time response to
an external force $R(t_w+\tau,t_w)$. For
temperatures higher than $T_c$, we have seen that $C$ and $R$ are actually
{\sc tti}, and furthermore that the {\sc fdt}
$R(\tau)=-\frac{1}{T} \Theta(\tau) \frac{\partial C(\tau)}{\partial \tau}$ is
obeyed. As noted in Section 3.2, one can then eliminate $R(\tau)$ and find an
equation for $C(\tau)$ which
 is precisely the schematic Mode-Coupling equation, with a kernel
 related to the correlation of the random potential.

 Hence, the physical
content of the (schematic) {\sc mct} is clear: it is a mean-field description
of a single
point in a static {\it quenched random} potential.
The important point is thus that {\sc mct} implicitly assumes the presence of
some
{\it quenched disorder} which should rather, as discussed above, be
`self-induced' by the dynamics itself.
In a sense it looks rather similar to the introduction of fiduciary models
discussed before.

\vskip 0.5cm
$\bullet$ Mode coupling at low temperatures
\vskip 0.5cm

Coming back to the general equations relating $C$ and $R$, one can now
postulate that they are the
correct generalisation of the schematic Mode Coupling equations for two time
quantities, and investigate the
`glass' phase $T <T_c$. The results of Section 3 are thus directly applicable.
In particular, the correlation and response function cease
to be functions of $\tau$ only. More precisely,
$C(t_w+\tau,t_w)$ can be
written as the sum of a stationary contribution $C_{\sc st}(\tau)$ which only
depends on
$\tau$, and an aging part which depends on the ratio $t_w/(t_w+\tau)$,
(or a generalisation thereof,
see Section \ref{onh(t)}): $C_{\sc ag}(t_w/(t_w+\tau))$. The expected shape of the correlation
function in the glass phase is
thus given in Fig. \ref{revfigClowT2}.

The same decomposition holds for the response function, and the aging parts of
$C$ and $R$ are related by an `anomalous' {\sc fdt},
where $T$ is replaced by an effective temperature $T/X$, with $X \leq
1$.

In more physical terms, this means that
for a finite waiting time $t_w$ after the quench below $T_c$, one expects that
the
susceptibility $\chi(\omega,t_w)$ still exhibits two peaks: a high frequency
$\beta$-peak very similar to the high temperature ($T>T_c$) one, and a low
frequency
$\alpha$-peak which reaches a maximum at a frequency $\omega_{\alpha} \simeq
1/t_w$, which thus progressively disappears as $t_w \to \infty$. An
interesting prediction of this low temperature extension of {\sc mct} is that
the high
frequency part of the aging $\alpha$-peak behaves as $(\omega t_w)^{-b}$, while
the low frequency `foot' of the $\beta$-peak behaves as $\omega^a$, with the
following relation between $a$, $b$, and $X$ (see (\ref{mctabLT})):
$
X \Gamma^2[1+b] / \Gamma[1+2b] = \Gamma^2[1-a]/\Gamma[1-2a].
$
This equation generalizes the well known {\sc mct} relation (\ref{mctabHT})
between $a$ and $b$ for
$T>T_c$, for which $X \equiv 1$. It would be extremely interesting to
try to test these predictions experimentally, taking care of the fact that the
above picture is only
valid insofar as $t_w$ is small compared to the relaxation time $\tau(T)$,
 such that `activated' effects might indeed be neglected.

\section{Conclusion. Where do we stand ?}
\label{Conclusion}

Let us now summarize some of the most important ideas developed in this review
and discuss
some open problems.

We have tried to show that aging effects are not spurious, irreproducible
artifacts of non equilibrium situations, but rather an unescapable feature of
systems characterized by a very large relaxation time, because
time-translational
invariance breaks down and the well known fluctuation-dissipation theorem has
to
 be modified in a non trivial way. However, the asymptotic aging regime where
all
times are large reveals some universal properties; in particular, the
`effective'
relaxation time becomes a time dependent notion and grows with the waiting
time.
The detailed investigation of these aging effects actually offer a
unique way to probe the phase-space structure of complex systems. From that
point
of view, a system in equilibrium is `dead'.

We have described
 several simple models where aging
can be described in detail: coarsening models, `trap' models and mean-field
spin-glass models: (which turn out to give dynamical equations in exact
correspondance with the `Mode-Coupling' description of supercooled liquids).
 Coarsening leads to aging in the correlation function but {\it not} in the
response
 function: the {\sc fdt} is most strongly violated in the aging regime.
This scenario can  thus  only include aging {\em in the response}
 of spin-glasses
 or polymer glasses  as a transient effect  even in an
infinite system
-- quite apart from the fact that it is not obvious
  what would actually {\it grow} with time in these systems.

Mean-field spin-glasses provide an interesting testing field. One generically
finds aging in the low temperature phase of these models, although two very
different categories of systems emerge. In the discontinuous case
(corresponding
in the high temperature phase to `model B' of {\sc mct}),
a dynamical temperature transition is found {\it above} the equilibrium
transition. Throughout
the low-temperature phase the asymptotic energy-density  arrived at after any
type of cooling is
higher than the equilibrium energy-density.
 The two-time plane breaks up into two sectors, which correspond to the
stationary
 and aging dynamical regimes.
On the contrary, for the  continuous case (such as the
Sherrington-Kirkpatrick model)
static and dynamical transition temperatures coincide, as do the asymptotic
out of equilibrium energy density and the equilibrium one.
The two-time behaviour
 is much more complicated, reflecting in some way the subtleties of Parisi's
 ultrametric equilibrium solution.

Interestingly, if one views mean-field glassy dynamics as the dynamics of a
point in a many
dimensional rugged phase-space, one finds that the basic mechanism for aging
 is germain to that of simple coarsening: because
of high dimensionality of phase-space the system starts
near the  border of a basin of attraction, and remains forever undecided about
where to
go.

It is of course crucial to know how these mean-field predictions are modified
when one goes
beyond mean-field and studies finite dimensional systems.
 From a theoretical point of view, the difficulty
 comes from the fact that activated processes, which are effectively absent in
 mean-field, come into play when the dimension is finite. For example,
the infinitely long-lived metastable states in mean-field acquire a finite
relaxation
time in finite dimension, through bubble nucleation. The dynamical transition,
 which corresponds in the language of supercooled liquids to the Mode-Coupling
 critical temperature, is thus smeared out. Similarly, a particle in a random
 potential in finite dimension reaches a local minimum after a finite time,
beyond
 which  thermal activation starts playing an important role;
 contributions to aging of a somewhat different nature ---
such as those described by the `trap' picture ---
then set in. The relation of these trap models to the aging dynamics
of real spin-glasses or other disordered systems such as pinned vortex lines,
 dislocations, domain walls, or polymers is however still rather tentative.

For the same reason (absence of activated effects), mean-field models are not
suited to
describe cooling rate dependent effects, which can be very large
in some disordered systems and in glass forming liquids.

Therefore, new theoretical ideas which would allow one to extend the
 previous approaches to finite dimensions (bearing in mind that activated
 effects cannot, in general, be accounted
for within perturbative schemes) are clearly desirable.
 Returning to the question of aging in the response,  the  {\sc
fdt}-violation  factor $X$ in finite dimension is particularly interesting:
 this factor --- that is related to effective temperatures in the system ---
is found to be non-trivial $(0 < X < 1)$ in the
 disordered mean-field models; what is the situation in finite dimensions?
 Some numerical results \cite{Frrie,binary} suggest that $X$ does
 indeed remain non trivial in finite dimensions, while arguments based on
coarsening pictures
would suggest that asymptotically, $X=0$. This might actually be
a clear-cut dynamical distinction between droplet like coarsening pictures and
 mean-field like pictures of real spin-glasses, which surely deserves more
efforts
 -- as stated above, spin-glasses, after all, {\it do} exhibit aging in
response functions.

We have also discussed rather at length the relation between glasses and
disordered
systems, noting that:

-- some models without disorder behave very much like disordered systems and
can
 actually be theoretically described as such, and

-- some approximation schemes for systems without disorder, but strongly
interacting, lead
to dynamical equations (i.e. the Mode-Coupling equations) which  correspond to
the hidden assumption of the existence of some quenched
disorder and
are actually exactly the
equations describing disordered mean-field spin-glasses.

Correspondingly, many of the ideas developed to describe aging in disordered
systems are
{\it de facto} also relevant for glasses. We have mentioned in particular how
aging
should manifest itself, within the Mode-Coupling scenario and for times smaller
than
the relaxation time as a waiting time dependent $\alpha$-peak. A detailed
analysis of
these aging regimes should enable one to distinguish, again, between mean-field
like
descriptions and activated, `trap' like, models; or perhaps understand how both
mechanisms
are blended.

Several questions still remain completely open. In particular, is there a
general
criterion allowing one to understand when a `complicated' system can be
described as
disordered ? Is this description only viable at finite times, where the
specificity of
the system at hand has not yet had time to reveal itself ? A possibility is
that this
time
scale can only be infinite for mean-field like models, such as the {\sc labs}
model or
the long-range Josephson array. Related to this is of course the lurking
question of
the very existence of a true glass transition in any short range system without
disorder.

The question of the existence of a true phase transition is however, after all,
not crucial
to understand the out-of-equilibrium properties, as it is
displayed in so many physical systems. In this respect, it is plausible that
mean-field models provide in general a much better starting point for the
finite time, dynamical
properties of real systems, than for its long time, equilibrium properties.

\vspace{2cm}

\noindent
\underline{Acknowledgements}
We have benefited from many useful discussions with
F. Alberici, A. Baldassarri, A. Barrat, R. Burioni, A. Comtet, D. S. Dean,
P. Doussineau, S. Franz, C. Godr\`eche, W. Kob, W. Krauth
A. Levelut, G. Lozano, C. Monthus, J. Hammann, L. Laloux, P. Le Doussal, R.
Monasson, M. Ocio, R. Orbach,
G. Parisi, L. Peliti, F. Ritort, H. Takayama, G. Tarjus, F. Thalmann, E.
Vincent,
M. A. Virasoro and H. Yoshino;
L. F. C. wishes to thank the SPEC at Saclay and the Laboratoire
de  Physique Th\'eorique des Liquides at Jussieu, Paris, where
part of this work was done, for their kind hospitality.


\begin{thebibliography}{99}
\bibitem{Phasespacemodels2}
Phase space models in the context of glasses have a
long history. For a review, see \cite{Science}.
\end{thebibliography}
\end{document}